\documentclass[journal,twocolumn]{IEEEtran}

\usepackage[switch]{lineno}

\usepackage{amsmath,amsthm,amssymb,multirow,tabularx,amsfonts,xspace,booktabs,comment,soul,caption,subcaption,afterpage,graphicx,algorithmic,textcomp,soul,mathtools,hyperref} 
\PassOptionsToPackage{hyphens}{url}
\usepackage[dvipsnames]{xcolor}
\usepackage{tikz}
\usetikzlibrary{patterns,hobby,backgrounds,calc,trees}
\usetikzlibrary{positioning,fit,calc,decorations.pathmorphing,patterns,shapes}
\usepackage{pgf}
\usepackage{pgfplots}
\pgfplotsset{compat=1.10}
\usepackage{colortbl}
\definecolor{Gray}{gray}{0.95}

\usepackage[dvipsnames]{xcolor}
\definecolor{manoscolour}{RGB}{153,51,255}

\usepackage[textsize=scriptsize, textwidth=16em]{todonotes}

\usepackage[nolist]{acronym}
\usepackage{url}

\newcommand{\util}{U}
\newcommand{\vulnerabilityset}{\mathcal{V}}
\newcommand{\vulnerabilityselected}{\mathcal{I}}
\newcommand{\overallreconcost}{S(\vulnerabilityselected)}
\newcommand{\reconcostset}{\mathcal{S}}
\newcommand{\payoffmatrix}{\Gamma}
\newcommand{\reconmatrix}{\Psi}
\newcommand{\game}{\mathcal{G}}

\newcommand{\low}{\ac{LIH}}
\newcommand{\high}{\ac{HIH}}
\newcommand{\abs}[1]{\left|#1\right|}
\newcommand{\set}[1]{\left\{#1\right\}}
\newcommand{\R}{\mathbb{R}}

\newcommand{\decept}{HoneyCar\xspace}
\renewcommand{\vec}{\mathbf}

\usepackage[normalem]{ulem}

\newif\ifNotes
\Notesfalse

\newif\ifExamples
\Examplestrue

\newtheorem{remark}[]{Remark}
\newtheorem{result}{Result}

\graphicspath{ {./figs/} }

\ifCLASSINFOpdf
\else
\fi
\begin{document}

%
\title{HoneyCar: A Framework to Configure Honeypot Vulnerabilities on the Internet of Vehicles}
%
%
%

\author{Sakshyam Panda,
        Stefan~Rass,~\IEEEmembership{Member,~IEEE,}
        Sotiris Moschoyiannis,~\IEEEmembership{Member,~IEEE,}\\
        Kaitai Liang,~\IEEEmembership{Member,~IEEE,}
        George Loukas,~\IEEEmembership{Member,~IEEE,}
        and~Emmanouil~Panaousis,~\IEEEmembership{Senior~Member,~IEEE}%
\thanks{S. Panda, G. Loukas, and E. Panaousis are with the School of Computing and Mathematical Sciences, University of Greenwich, London SE10 9LS, UK, 
e-mail: \{s.panda, g.loukas, e.panaousis\}@greenwich.ac.uk.}
\thanks{S. Rass is with the LIT Secure and Correct Systems Lab, Johannes Kepler University Linz, Linz, Austria, 
e-mail: stefan.rass@jku.at.}
\thanks{S. Moschoyiannis is with the Department of Computer Science, University of Surrey, Guildford GU2 7XH, UK,
e-mail: s.moschoyiannis@surrey.ac.uk.}
\thanks{K. Liang is with the Cyber Security Group, Delft University of Technology, Delft, Netherlands, 
e-mail: Kaitai.Liang@tudelft.nl.}
\thanks{``This work has been submitted to the IEEE for possible publication. Copyright may be transferred without notice, after which this version may no longer be accessible.''}
}

%
%

%

\maketitle

\begin{abstract}
The Internet of Vehicles (IoV), whereby interconnected vehicles communicate with each other and with road infrastructure on a common network, has promising socio-economic benefits but also poses new cyber-physical threats. 
Data on vehicular attackers can be realistically gathered through cyber threat intelligence using systems like honeypots. 
Admittedly, configuring honeypots introduces a trade-off between the level of honeypot-attacker interactions and any incurred overheads and costs for implementing and monitoring these honeypots. 
We argue that effective deception can be achieved through strategically configuring the honeypots to represent components of the IoV and engage attackers to collect cyber threat intelligence. 
In this paper, we present HoneyCar, a novel decision support framework for honeypot deception in IoV.
HoneyCar builds upon a repository of known vulnerabilities of the autonomous and connected vehicles found in the Common Vulnerabilities and Exposure (CVE) data within the National Vulnerability Database (NVD) to compute optimal honeypot configuration strategies. 
By taking a game-theoretic approach, we model the adversarial interaction as a repeated imperfect-information zero-sum game in which the IoV network administrator chooses a set of vulnerabilities to offer in a honeypot and a strategic attacker chooses a vulnerability of the IoV to exploit under uncertainty. 
Our investigation is substantiated by examining two different versions of the game, with and without the re-configuration cost to empower the network administrator to determine optimal honeypot configurations. 
We evaluate HoneyCar in a realistic use case to support decision makers with determining optimal honeypot configuration strategies for strategic deployment in IoV. 
\end{abstract}




\IEEEpeerreviewmaketitle
\frenchspacing

%
%
\section{Introduction}
The Internet of Vehicles (IoV) is revolutionising the transportation industry with increasing number of connected vehicles. 
The aim is to achieve an integrated intelligent transportation system with advanced traffic management to ensure road safety, avoid road accidents and improve driving experiences.
IoV is a distributed network that utilises the data gathered by vehicles to improve safety on roads and allow interaction of vehicles with other heterogeneous networks through smart Vehicle to Everything (V2X) communication \cite{kaiwartya2016internet}. IoV has evolved from Vehicular ad hoc networks (VANETs) and is expected to eventually evolve into the Internet of Autonomous Vehicles \cite{gerla2014internet}. 

Although connected vehicles bring benefits for the passenger experience and road safety, they have also introduced more opportunities for cyber attackers to compromise the vehicles endangering passengers and pedestrians lives.
Additionally, the growing prevalence of cyber incidents has resulted in a strong demand for security in IoV solutions to ensure cost-effective, scalable and robust services that conform to legal requirements and adequately confront cyber-physical threats, where a cyber security breach may lead to a privacy breach or adverse physical impact \cite{CPA15}. 
Steady functioning of IoV requires integration of many different technologies, services and standards \cite{cheng2015routing}. 
As vehicles become more connected to their vulnerable and unprotected external environment, the number of attack possibilities and the risk of a vulnerability being exploited greatly increases.
IoV thus needs to address numerous security issues including threats concerning the Internet of Things (IoT) but also new threats specific to connected vehicular networks.
Cyber deception and more specifically honeypots have been studied within the field of vehicular networks, e.g., \cite{patel2017honeypot,verendel2008approach,gantsou2014toward,schmitzstrategy}.

In this paper we propose a game-theoretic framework to address the strategic interaction between an agent that chooses a set of vulnerabilities to implement in a honeypot and a competing agent that selects a vulnerability for exploitation of the targeted IoV. 
Numerous papers have used game theory to model deception \cite{la2016deceptive,zhu2021survey,la2016game} and other security challenges \cite{manshaei2013game,do2017game}.
In particular, using zero-sum games, according to which the loss of an agent is the gain of the other agent is the most acceptable approach in the literature, e.g., \cite{panaousis2017game,pibil2012game}.

Based on the target location of the attacker, attacks on IoV are classified into two groups \cite{sakiz2017survey}: (i) Inter-vehicle attacks where compromised vehicles could misbehave or falsify data sent to other vehicles leading to drastic impact on human lives, energy and money; and (ii) Intra-vehicle attacks that involve deceiving a sensor or a system resulting in damage to the vehicle and the environment. 
The cyber threats that affect IoV can range from the application layer all the way down to the physical layer. 
Akin to the significant growth and value of the autonomous and connected vehicles, there is a growing concern about the cyber security and privacy preservation within these systems. 
Smart vehicles are not only exposed to specific risk in this context, but some of the risks may involve risk to human life \cite{ring2015connected}. 
One of the current cyber security efforts with respect to vehicles can be found in the regulations developed within
the United Nations Economic Commission for Europe (UNECE). 
In particular, the new UNECE regulations\footnote{UN Regulation No. 155 - Cyber security and cyber security management system, \url{https://unece.org/transport/documents/2021/03/standards/unregulation-no-155-cyber-security-and-cyber-security}} highlight theimportance of detection and mitigation of cyber incidents.
Threat detection, in turn, relies on cyber threat intelligence and defensive deception as effective means to collect information on adversaries intents and actions \cite{zhu2021survey}. 

Identifying risk mitigation actions is one of the main challenges of security decision-makers in all sectors. 
As IoV inherit many of the characteristics of regular computer networks, most of the countermeasures for computer networks are also applicable to IoV. 
Some of the proposed countermeasures include threat modelling of security risks in IoV \cite{lam2021ant}, intrusion detection systems for protection against internal and external attacks \cite{aloqaily2019intrusion}, secure routing protocols \cite{rivas2011security,emara2015evaluation} and privacy mechanisms \cite{cheng2015routing}, encryption key management \cite{wazid2019akm} and honeypots to collect adversarial information in IoV \cite{sharma2018survey}. Additionally, security frameworks and standards such as the Cybersecurity Framework for Critical Infrastructure \cite{barrett2018framework}
proposed by the National Institute of Standards and Technology (NIST) could be incorporated to strengthen the security of IoV.

In IoV, the authorisation and the communication modules which handle the most critical security features are often targeted by attackers leading to serious security, privacy, and physical impacts. 
Honeypots can significantly contribute in protecting these systems by absorbing damage and collecting information about attackers' intents and actions. 
The knowledge from honeypots can be leveraged to identify and close security gaps, implement measures to avoid attacks on vehicles and infrastructure, support intrusion detection and prevention systems, and to realise a wide range of attacks and threats to IoV. 
Even so, it is important to know how attackers gain access and proceed in the network and how the vehicles and its connected environment behave in case of such attacks. 
Honeypots in general have advantages over standard intrusion detection systems as any connection to the honeypot can be considered suspicious thus being useful to the network administrator in identifying threats and malicious activities.

However, a key challenge in designing honeypots is convincing attackers that these are real systems \cite{rowe2006measuring}.
As honeypots do not involve regular users and usage patterns, it is a challenge to design convincing honeypots. 
As highlighted in existing literature \cite{sharma2018survey}, the problem can be addressed by strategically configuring honeypots such that they resemble components of the IoV network. 
Further, as there is always a limitation to the amount of resources to be invested in cybersecurity controls, there is a need to strategically deploy them to improve security as well as have a positive Return On Security Investment (ROSI). 
To assist with security decisions, existing literature proposes several cost-benefit approaches for selecting optimal set of countermeasures against cyber attacks \cite{fielder2016decision,nespoli2017optimal} leading to optimal cybersecurity investment plans \cite{gordon2002economics,chronopoulos2017options}. 

This paper proposes a novel decision support framework, called \decept, that addresses the challenges in designing convincing honeypots. 
Our research is motivated by the fact that an effective cybersecurity strategy must consider the advancing threat landscape, which can only be achieved by understanding attackers' actions.
\decept allows the decision maker to compute optimal strategies for active re-configuration of honeypots based on a set of available vulnerabilities, an available budget and the expected benefit, in terms of attacker engagement time, acquired through each available configuration action. 
Note that, we only consider long-distance communication mode in IoV which includes most of the critical connections concerning security \cite{sharma2018survey}; in particular, V2X which supports all the communication between a vehicle and its environment whether it is another vehicle, roadside infrastructure, or the cloud. 

\decept consists of two models: (i) a formal model of investing in honeypot deception; and (ii) a game-theoretic model where the decision maker (called the Defender) configures a honeypot in IoV while the adversary (called the Attacker) attempts to exploit vulnerabilities of the observed systems in the network. 
The game-theoretic model presents the strategic use of honeypots by modelling the interaction between the Defender and the Attacker using a repeated imperfect-information zero-sum game. 

To sum up, this work makes the following contributions:
\begin{itemize}
    \item It presents a novel decision support framework called \decept to examine practical (cost effective) honeypot investment decisions.    
    
    \item It presents a novel game-theoretic model demonstrating the strategic use of honeypot through cost effective selection and configuration of honeypot based on a given budget. 
    
    \item Through a case study using vulnerabilities related to autonomous and connected cars in the Common Vulnerabilities and Exposure (CVE) list of records\footnote{\url{https://cve.mitre.org/}}  with the associated Common Vulnerability Scoring System (CVSS) metrics\footnote{\url{https://www.first.org/cvss/}}, it demonstrates how \decept can assist with honeypot configuration and investment decisions satisfying various objectives of using honeypots in IoV.  
    
    \item It highlights the significance of re-configuration cost in making well-informed cybersecurity investment plans.
\end{itemize}

The rest of this paper is structured as follows. 
Section \ref{sec:relatedwork} presents the related work on the use of honeypots in connected vehicular networks. 
It also presents an overview on the application of game theory to honeypot deception to secure networks and highlights how this work extends the current literature.
Section \ref{sec:model} introduces the honeypot deception model together with the decisions of the Defender and the game-theoretic model for optimal configuration of honeypots. 
Section \ref{sec:analysis} presents the mathematical analysis for computing optimal honeypot configurations to support the investment decisions.
Section \ref{sec:evaluation} presents a case study using the known vulnerabilities related to the autonomous and connected vehicles to evaluate \decept.
Finally, Section \ref{sec:conclusion} concludes this paper by discussing the limitations of the proposed method and potential future work.

%
%

\section{Related Work}\label{sec:relatedwork}

IoV have many open security problems \cite{hasrouny2017vanet} from identification and communication to standardisation \cite{vermesan2011internet}. 
These threats challenge the privacy and security in IoV \cite{qu2015security}.
However, research has preliminary focused on identifying vulnerabilities in connected and autonomous vehicles and analysing the potential impact of successful exploitation while proposing some mitigation measures \cite{hasrouny2017vanet,petit2014potential}. 
Successful cyber attacks on IoV networks have been documented including those on security keys used in electronic control units, wireless key fobs, tyre-pressure monitoring systems, inertial measuring units, braking system, and more \cite{parkinson2017cyber,zhang2014defending,miller2015remote}. 


Deception is used for both defensive and offensive interactions. 
Researchers have developed various defences against offensive deception in connected vehicular networks such as sybil attacks \cite{li2021rted}, false positioning attacks \cite{singh2019machine}, illusion attacks \cite{lo2007illusion} and topology poisoning attacks \cite{wang2020topology}.
Honeypots have been widely implemented in network security for defensive deception \cite{carroll2011game,boumkheld2019honeypot}, moving target defence \cite{zhu2013game} and against advanced persistent threats \cite{tian2021honeypot}.  
A broad range of attacks on connected and autonomous vehicles could be detected by using honeypots. 
These attacks could include random attacks on the vehicles and connected infrastructure as well as targeted attacks on IoV.

\subsection{Honeypots for Vehicles}

Honeypots have been used to identify vehicles behaving selfishly to preserve their resources \cite{patel2017honeypot}, collect adversarial information \cite{verendel2008approach}, and support intrusion detection systems for VANETs \cite{sharma2018survey}.
\cite{gantsou2014toward} highlighted the application of honeypots in VANETs by presenting how in-vehicle honeypots and road side unit honeypots could be used to support the security and enhance the performance of VANETs. 
Verendel et al. \cite{verendel2008approach} proposed three in-vehicle honeypot models demonstrating the use of honeypots in IoV to gather adversarial information.
Besides, \cite{schmitzstrategy} proposed the use of a low-interaction honeypot together with a high-interaction honeypot to collect threat and vulnerabilities data in connected cars. 
\cite{you2020scalable} proposed a physical honeypot with cross-network access, one-to-many mechanisms and high interaction capabilities replicating a programmable logic controller to gather large-scale attack data.

To detect blackhole attacks in wireless mesh networks such as VANETs and IoV, \cite{prathapani2009intelligent} and \cite{babu2016novel} used honeypots to generate dummy route request packets to lure and isolate blackhole attackers. 
Patel and Jhaveri \cite{patel2017honeypot}, on the other hand, used honeypot-based detection technique to identify vehicles in VANETs behaving selfishly to save their resources.
Our research extends the frontier of the application of honeypots to secure autonomous and connected vehicles.
It focuses on the strategic re-orientation of honeypots to deceive attackers and collect threat data in IoV.


\subsection{Deception}

In the context of deceiving attackers by implementing honeypots, Garg and Grosu \cite{garg2007deception} investigated the allocation of honeypots in a network as a signalling game in which the Defender implements $k$ honeypots out of $n$ possible host systems.
The goal was to deceive attackers by placing several honeypots in the network. When attackers cannot determine the type of systems due to deployed deception, they either postpone the attack or spend additional resources to determine the identity of a system.
Carroll and Grosu \cite{carroll2011game} studied a similar signalling game in which the Defender can disguise a real system as a honeypot and a honeypot as a real system. 
The authors analysed both pooling and hybrid \emph{Perfect Bayesian Nash Equilibria} while demonstrating the in-feasibility of separating equilibria. 
Ceker et al. \cite{cceker2016deception} build on \cite{carroll2011game} to design a honeypot-based deception method to mitigate distributed denial-of-service attacks. 
The Defender can either choose a system to be a real system or a honeypot and the Attacker can either observe, attack or retreat. 

On the other hand, P{\'\i}bil et al. \cite{pibil2012game} and Kiekintveld et al. \cite{kiekintveld2015game} investigated ways of designing honeypot systems to optimise the probability of the Attacker choosing to attack the honeypot rather than the real system. 
The authors modelled the Defender-Attacker interaction using a zero-sum game. 
Their model classified systems into different levels of importance with assigned utilities for the Attacker where the Defender played to minimise the Attacker's expected value.
La et al. \cite{la2016deceptive} analysed an IoT network along with a honeypot to defend systems from attacks and extended the analysis from a single-shot game to a repetitive game considering the deceptive aspects of the players.  
While most of the aforementioned works investigated the deceptive use of honeypots to minimise the probabilities of attacks on the real system, the ways of engaging the attackers remained understudied. 
\decept closes this gap by looking into the use of honeypot to deceive adversaries with the explicit goal of optimal re-configuration of the honeypot to increase Attackers' engagement, and thereby enhancing the information gathered on adversarial behaviour.

There are relatively fewer papers that use game theory to optimise the information learned about attackers through multi-round games. 
Zhuang et al. \cite{zhuang2010modeling} proposed a multi-round signalling game to investigate strategies of deception and resource allocation. 
In each round of the game, Defender decides on investing in either short term expense or long term capital investments in defence. 
To deceive the Attacker, Defender then chooses how much of the investment information to reveal as a signal. The Attacker on observing the signal decides whether to attack or not. Durkota et al. \cite{durkota2015optimal} investigated a similar multi-round interactions as a Stackelberg game. 
The Defender plays first by placing honeypots to harden the network. 
The Attacker with knowledge on the number of honeypots but not their identities follows the Defender. 
Attack graphs are used to represent Attacker's multi-round strategies and develop network hardening techniques to enhance security. 
Unlike most game-theoretic models of deception which are either static games or single-shot games, we study the interaction between the Defender and the Attacker as a multi-round game. 
Also, concerning the structure of the game, the work closest to ours is that of Durkota et al. \cite{durkota2015optimal} for regular computer networks. 
\decept improves on this by introducing re-configuration (similar to hardening operation) of honeypot in each round of the game to engage attackers, rather than just placing honeypots to harden IoV network. 
Finally, \decept uses known vulnerabilities from the CVE records along with the Common Vulnerability Scoring System (CVSS) metrics to provide actionable recommendations providing decision support to the IoV network administrator as will be shown in Section \ref{sec:evaluation}.

%
%
\section{System Model}\label{sec:model}
We assume that a service provider (e.g., management service or network provider), henceforth called the Defender, decides to invest in honeypots to dissuade adversaries from critical infrastructure and to capture adversarial activities. 
An adversarial entity is referred to as the Attacker.  
To achieve this, the Defender must design their honeypots such that they can successfully deceive attackers into considering these honeypots as systems of importance while the investment in honeypots being economical.
Successful deception may lead to identification of potential attributes of intrusion and techniques of the Attacker. 
This cyber threat information can enable the Defender to strengthen their defences to prevent attacks on real systems by identifying countermeasures against the observed security gaps, augmenting intrusion detection signatures or anomaly-based rules based on the observed behavioural patterns of the Attacker. 
The information can also be used to support forensic investigations \cite{nisioti2021data} of cyber security breaches.

\subsection{System Model}

The Defender is often confronted with the challenging task of making security investment decisions with a limited security budget and a strict reaction time frame. 
Key questions the Defender faces are: (i) \textit{whether to invest in cyber deception} and (ii) \textit{how this deception must be implemented and delivered}.
To support these decisions, we investigate a scenario where the Defender has an allocated investment budget and this budget can be determined by another cyber investment decision-making methodology such as \cite{fielder2016decision,panda2020optimizing,fielder2018risk}. 
The computation of this amount as well as its size is out of the scope of this paper and the initial question is whether to implement a honeypot or not given this budget.

Let $B$ be the required budget, i.e., cost, for covering the implementation and maintenance of a honeypot. 
We denote by $L$ the expected loss caused by a successful breach on a system in the IoV network. 
Adding a honeypot in the network would be beneficial when $L$ is greater than $B$ as investing in deception is affordable and less expensive than the expected loss caused from cyber incidents. Further, the information gathered from the honeypot could be used to reduce this expected loss.
Likewise, the Defender would prefer not to implement a honeypot when $L<B$ as the implementation cost would exceed its reward leading to a negative Return on Security Investment (ROSI).

When investing is a preferable choice, the next step is to identify what type of honeypot to implement i.e., either a \ac{LIH} or a \ac{HIH}. 
Each honeypot type has an implementation cost, expressed as $C_l$ and $C_h$ for $\low$ and $\high$, respectively, and a learning rate ($\lambda$). 
The implementation cost is exogenous and depends on the resources required to implement and maintain it. 
For example, $\low$s are easy to deploy and maintain, usually hosted on a virtual machine, and offer limited services such as Internet protocols and network services without any interactions with the operating system.    
The limited interaction enables them to minimise the risk of exploitation by containing the activities of the Attacker within the deception environment. 
These low-level interactions only capture limited information on the Attacker's activities, thus leading to a lower learning rate.  
On the other hand, $\high$s provide greater levels of interaction such as interactions with a real/virtual operating system. 
These additional functionalities introduce complexities in implementing and maintaining them \cite{fan2017enabling}, besides the higher risk of breaching the real systems through them. The learning rate for $\high$ is higher as they can capture more activities of the Attacker contributing to better cyber threat intelligence. 
Table \ref{tab:list_symbols} presents the list of symbols used in this paper. 

\begin{table}[t]
\centering
\small
\renewcommand*{\arraystretch}{1.1}
\resizebox{\columnwidth}{!}{\begin{tabular}{|c|l|}
  \hline
  \textbf{Symbol} & \textbf{Interpretation} \\
  \hline
  \multicolumn{2}{|c|}{Constants and Functions}\\
  \hline
  \rowcolor{Gray}$B$    		& budget to invest in honeypot \\
  $C_h$                         & implementation cost of $\high$ \\
  \rowcolor{Gray}$C_l$  		& implementation cost of $\low$ \\
  $\vulnerabilityset$           & set of available vulnerabilities \\
  \rowcolor{Gray}$\vulnerabilityselected$ & set of offered vulnerabilities ($\vulnerabilityselected \subseteq \vulnerabilityset$)\\
  $L$                           & expected loss without honeypot \\
  \rowcolor{Gray}$\reconcostset(\vulnerabilityselected)$ & re-configuration cost for $\vulnerabilityselected$ \\
  \hline
  \multicolumn{2}{|c|}{Game Variables}\\
  \hline
  \rowcolor{Gray}$B_{res}$      & residual budget after implementing honeypot \\
  $s_j^+$                       & cost of adding vulnerability $v_j \in \vulnerabilityselected$ \\
  \rowcolor{Gray}$s_j^-$        & cost of patching vulnerability $v_j \in \vulnerabilityselected$ \\
  $c(\vulnerabilityselected)$   & honeypot monitoring cost when offering $\vulnerabilityselected$\\
  \rowcolor{Gray}$g(\vulnerabilityselected)$ & gained cyber threat intelligence when offering $\vulnerabilityselected$\\
  $p_j$                         & probability of offering vulnerability $v_j \in \vulnerabilityselected$\\
  \rowcolor{Gray}$t_j$          & time to exploit vulnerability $v_j \in \vulnerabilityselected$\\
  $\util_h$                     & expected game utility of the Defender with $\high$\\
  \rowcolor{Gray}$\util_l$      & expected game utility of the Defender with $\low$\\
  $\nu(\vulnerabilityselected)$ & optimum value obtained from optimisation \\
  \rowcolor{Gray} $\alpha$      & average monitoring cost factor \\
  $\beta$                       & optimisation smoothing factor \\
  \rowcolor{Gray} $\lambda$     & honeypot learning rate \\     
  \hline
\end{tabular}}
\caption{List of Symbols}
\label{tab:list_symbols}
\end{table}

We also assume that each honeypot type can only support vulnerabilities of certain complexities.
Let $\vulnerabilityselected \subseteq \vulnerabilityset$ denote the set of vulnerabilities to be offered in a honeypot. 
The complexity to exploit a vulnerability, i.e., low, medium, or high, is extracted from \ac{CVSS} and is used to differentiate vulnerabilities that can be offered through $\low$ and $\high$.
We assume vulnerabilities with ``high'' complexity can only be offered through $\high$. 
The Defender has to strategically offer the vulnerabilities to engage the Attacker, which could involve re-configuring the honeypot multiple times. The re-configuration activities can be achieved using honey-patches \cite{araujo2014patches}, also known as ghost-patches \cite{avery2017ghost}. 
A honey-patch can function similar to a regular patch and has two components: (i) a traditional patch for the known software vulnerability, and (ii) additional code to to create fake vulnerabilities to mislead the Attacker \cite{araujo2014patches}.
Honey-patches can lead the Attacker to the decoy providing a false sense of success.
However, the Defender has to bear costs to perform the re-configuration and maintenance of the honeypot. 
We refer to these costs as the re-configuration cost $\overallreconcost$. 
The re-configuration action is analysed using a game-theoretic framework based on a repeated game, as discussed in the following section.
Besides, respecting the available budget $B$, the Defender's security investment decisions are determined by:
\begin{itemize}
    \item the expected loss $L$, without a honeypot, due to a breach;
    \item the choice of honeypot type to implement $\in\{l,h\}$;
    \item the set of vulnerabilities to offer in a honeypot, $\vulnerabilityselected = \{v_1, v_2, \cdots, v_n\} \subseteq \vulnerabilityset$, and their associated exploitation times $\set{t_1, t_2, \cdots, t_n}$. 
    \item the cost of re-configuration $\overallreconcost = \{(s^+_1,s^-_1), (s^+_2,s^-_2),\\ \cdots, (s^+_n,s^-_n) \}$, where $s^+_j$ and $s^-_j$ are costs (in time) for deliberately \textit{opening} (indicated by ``+'' superscript) or knowingly \textit{patching} (`-'' superscript) vulnerability $v_j \in \vulnerabilityselected$, respectively. 
\end{itemize}

\subsection{Game Model}

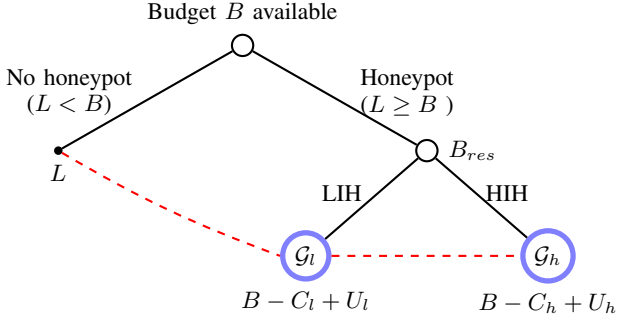
\begin{figure}[t]
\centering
\begin{tikzpicture}[scale=1.4,thick,-,text centered,level distance=1cm,
  font=\small,
  solidnode/.style={circle,draw,inner sep=1mm},
  hollownode/.style={circle,draw=blue!50,inner sep=.8mm,fill=white,line width=2},
  end/.style = {circle,draw,inner sep=0.3mm, fill=black},
  level 1/.style={sibling distance=3.5cm},
  level 2/.style={sibling distance=2.3cm}
  ]

\node (root) [solidnode, label=north:{Budget $B$ available}] {}
  child{
    node(1)[end, label=south:{$L$}] {}
    edge from parent node[left, xshift=-2, yshift=3, align=center]{No honeypot \\ ($L < B$)}
  }
  child{
    node[solidnode, label=east:{$B_{res}$}] {}
      child{
        node(2)[hollownode, label=south:{$B - C_l + \util_l$}] {$\game_l$}
        edge from parent node[left, xshift=0, yshift=2, align=center]{\low}
      }
      child{
        node(3)[hollownode, label=south:{$B - C_h + \util_h$}] {$\game_h$}
        edge from parent node[right, xshift=-2, yshift=2, align=center]{\high}
      }
    edge from parent node[right, xshift=6, yshift=2, align=center]{Honeypot \\ ($L \ge B$ )}
  };

\draw[red, dashed, thick] (1) to[bend right=5] (2.west);
\draw[red, dashed, thick] (2) to (3);
\end{tikzpicture}
\caption{Defender's decision tree with sub-games $\game_l$ and $\game_h$}
\label{fig_honeypot_configuration_game}
\end{figure}

The available choice for the Defender is to either implement $\low$ or $\high$. 
Each honeypot type leads to a distinct strategic game, played between the Defender and the Attacker, as illustrated in Figure \ref{fig_honeypot_configuration_game}. 
We refer to these games as Honeypot Configuration Games (HCG) represented as $\game_l$ and $\game_h$ (related to $\low$ and $\high$, respectively).
We utilise game theory \cite{fudenberg_game_1991} which studies optimal decisions involving multiple decision makers (also referred to as agents or players), including adversarial settings where two or more players have opposing goals, to support the optimal decision.

We model the HCGs as a \textit{repeated imperfect-information zero-sum two-player game}. 
Security games involve interaction between players with exactly opposite goals making zero-sum games the best fit to model them \cite{do2017game}. 
Zero-sum games, in particular, capture the worst-case scenarios for the Defender, which is to face the Attacker who is after the most valuable assets.
The key motivation behind using zero-sum games is that they can provide desirable solutions against any opponent and not just against rational opponents \cite{pibil2012game}. 
On some occasions, this may mean that the Defender spends more resources in particular when the Attacker is non-strategic or naive. 
However, in the absence of complete knowledge about the Attacker's incentives, it is reasonable to model the proposed investment and honeypot configuration decisions as a zero-sum game to achieve robustness against strategic adversaries, which most often are mentioned in the literature as Advanced Persistent Threats (APTs) \cite{pitropakis2018enhanced}.

Repeated games refer to games that have multiple independent rounds. 
In each round of HCG, the Defender re-configures the honeypot. 
Re-configuring is a hardening operation which involves replacing an observed exploited vulnerability with another from the $\vulnerabilityselected$. 
The replacement choice aims to convince the Attacker that the honeypot is a valuable system. 
The important aspect of this modelling decision is that \decept considers not only the costs for implementing and maintaining the honeypots but also the cost for re-configuring the honeypot. 
This comes into the model via the cost variables $s_i^+$ and $s_i^-$ to offer or revoke some vulnerability $v_i$ in a new configuration.

In each HCG, the Defender chooses a set of vulnerabilities $\vulnerabilityselected \subseteq \vulnerabilityset$ to offer. 
The offered vulnerabilities can be observed by the Attacker during system reconnaissance, who in response chooses a vulnerability from the offered ones to exploit. 
Having a honeypot in the network makes it difficult for the Attacker to gain the exact configuration of the network as he cannot distinguish a real system and a honeypot with certainty.
The Attacker also has no information regarding the kind of the system he is targeting and the value of the asset he will have access to by exploiting the vulnerability. 
Furthermore, we make the following assumptions about the Attacker: (i) he needs to compromise at least one vulnerability to mount an attack on the IoV network; and (ii) he is aware of the possibility of a honeypot (decoy system) and plays the best strategy, that is, targets a vulnerability that maximises the chances of successfully breaching the targeted system.

$\low$ is a cheaper option than $\high$, but it is more likely to be recognised by the Attacker leaving the Defender with a lower reward.
The Defender, therefore, has to undertake some cost-benefit analysis to identify the best type of honeypot for defending the IoV network. 
The choices of the players determine the game utility which is the outcome of the cost-benefit analysis performed when these strategies are played.
As illustrated in Figure \ref{fig_honeypot_configuration_game}, the choice of a honeypot type to implement depends on the residual security budget defined as
\begin{align} \label{eq_B_res}
   B_{res} = \text{max } \Big\{B - C_l + \util_l ~, B - C_h + \util_h  \Big\}
\end{align}
where $\util_l$ and $\util_h$ are the Defender's expected game utilities from $\game_l$ and $\game_h$, respectively. 
These utilities correspond to the payoffs of the Defender for implementing a honeypot type with a set of vulnerabilities ($\vulnerabilityselected$) to be exploited by the Attacker. 
In each of these games, the Defender selects $\vulnerabilityselected \in \vulnerabilityset$, which is different for $\low$ and $\high$. 
It is evident from Figure \ref{fig_honeypot_configuration_game} that the selection of a honeypot type and the vulnerabilities to offer would alter $B_{res}$.

\begin{remark}
    The expected game utility is a function of the gained cyber threat intelligence and the cost of monitoring the offered vulnerabilities.
\end{remark}

\subsection{Game iterations and utilities}

In HCG, the Defender configures the honeypot with a selection of $m$ out of $n=\abs{\vulnerabilityset}$ possible vulnerabilities to let the Attacker mount any available exploit on the offered $m$ weaknesses denoted by $\vulnerabilityselected=\{v_1, \ldots, v_m\} \subseteq \vulnerabilityset$. 
The Attacker would anticipate that the Defender will eventually discover the vulnerability expecting the weakness to be patched as in a real system. 
But other vulnerabilities will remain, and new vulnerabilities are opened up with an attempt to retain the attractiveness of the honeypot for the Attacker. 
This is exactly the dynamics that HCG implements. 
The games proceed with an assumption that in each iteration the Attacker exploits only one vulnerability from the offered ones and the Defender re-configures the honeypot by patching the exploited vulnerability to demonstrate activity as she would on a real system. 
The Defender further offers a new vulnerability $v_t$ from the set of remaining vulnerabilities as an attempt to keep the Attacker interested in the honeypot. 

\begin{figure}[th]
\centering
    \begin{subfigure}[b]{1\columnwidth}
        \begin{tikzpicture}[scale=1.7,thick,-,text centered,level distance=1.2cm, font=\small,
          solidnode/.style={circle,draw,inner sep=1mm},
          hollownode/.style={circle,draw,inner sep=.8mm,fill=white,line width=2},
          end/.style = {circle,draw,inner sep=0.3mm, fill=black},
          level 1/.style={sibling distance=1cm},
          level 2/.style={sibling distance=1cm}
        ]
            \node (root) [solidnode, label=north:{Defender}] {}
              child{
                node(1)[end, label=south:{$\util_1$}] {}
                edge from parent node[left, xshift=-2, yshift=3]{$v_1$}
              }
              child{
                node(2)[end, label=south:{$\util_2$}] {}
                edge from parent node[left, xshift=2, yshift=3]{$v_2$}
              }
              child{
                node(i)[end, label=south:{$\util_t$}] {}
                edge from parent node[left, xshift=-2, yshift=3]{$v_t$}
              }
              child{
                node(n)[end, label=south:{$\util_m$}] {}
                edge from parent node[right, xshift=-2, yshift=3]{$v_m$}
              };

            \node [left = 0.25cm of 1, yshift = -0.25cm, text width = 1.5cm]
                                         () {Defender's expected payoff};
            \path (2) -- node[auto=false]{\ldots} (i);
            \path (i) -- node[auto=false]{\ldots} (n);
        \end{tikzpicture}
        \caption{A round of HCG.}
        \label{fig:HCG_offered_vul}
    \end{subfigure}
    
    \begin{subfigure}[b]{0.7\columnwidth}
    \centering
        \begin{tikzpicture}[scale=1.7,thick,-,text centered,level distance=1.2cm, font=\small,
          solidnode/.style={circle,draw,inner sep=1mm},
          hollownode/.style={circle,draw,inner sep=.8mm,fill=white,line width=2},
          end/.style = {circle,draw,inner sep=0.3mm, fill=black},
          level 1/.style={sibling distance=1cm},
          level 2/.style={sibling distance=1cm}
        ]
            \node (root) [solidnode, label=north:{Defender}] {}
              child{
                node(1)[end, label=south:{$\util_1$}] {}
                edge from parent node[left, xshift=-2, yshift=3]{$v_1$}
              }
              child{
                node(i)[end] {}
                edge from parent node[left, xshift=-2, yshift=3]{$v_t$}
              }
              child{
                node(j)[end, label=south:{$\util_j$}] {}
                edge from parent node[right, xshift=-5, yshift=3]{$v_j$}
              }
              child{
                node(n)[end, label=south:{$\util_m$}] {}
                edge from parent node[right, xshift=-2, yshift=3]{$v_m$}
              };

            \path (j) -- node[auto=false]{\ldots} (n);
            \draw[decoration = {zigzag,segment length = 3mm, amplitude = 1mm},decorate, very thin] (i) -- (root);
        \end{tikzpicture}
        \caption{A later round with changed vulnerabilities.} 
        \label{fig:HCG_patch_new_vul}
    \end{subfigure}
    \caption{The dynamic of HCG: (i) In a round of HCG, the Defender offers a set of vulnerabilities to be exploited by the Attacker; (ii) In the next round of HCG, the Defender patches a vulnerability $v_t$ and offers a new one $v_j$ from $\vulnerabilityset$.}
\end{figure}
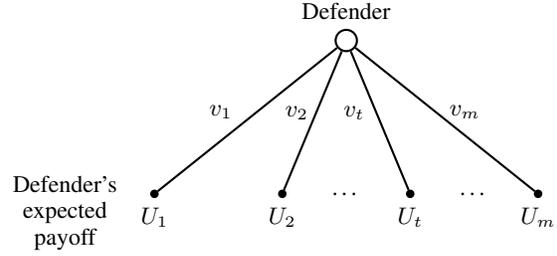
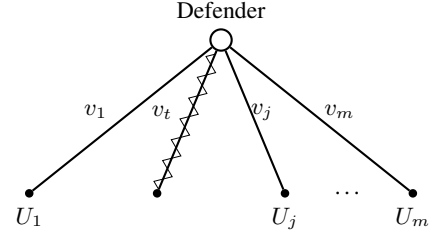

Each vulnerability has its own cost and benefits based on the time required to exploit it and the observed activities of the Attacker during the interaction with the vulnerability. 
For reference, Figure \ref{fig:HCG_offered_vul} and Figure \ref{fig:HCG_patch_new_vul} illustrate the players' interaction in HCG.
The strategic choices of the players are the following:
\begin{itemize}
  \item the Defender chooses an $m$-element subset $\vulnerabilityselected \subseteq \vulnerabilityset$, making the strategy space having cardinality $\binom{n}{m}$.
  \item the Attacker chooses a vulnerability $v_j \in \vulnerabilityselected$ to exploit. This choice can be motivated based on several criteria such as time, resources and skills.
\end{itemize}

\begin{remark}\label{rem:adv-profiling}
    Investigating adversarial profiles is beyond the scope of this paper and for simplicity, yet realistic, we consider that the Attacker prefers the ``easiest'' of all $v_t \in \vulnerabilityselected$ to break into a targeted system. Likewise, we do not further study sequential combinations of several exploits, and focus our analysis on a single exploitation trial, corresponding to a single round of our game. Repeated attempts as occur in practice then manifest as rounds of our repeated game.
\end{remark}

The Attacker aims at minimising the time spent to exploit $v_t$ to decrease the engagement time with a targeted system. 
Furthermore, intelligent attackers always aim to compromise a targeted system while remaining undetected. 
The execution of such a stealthy attack allows the attacker to compromise the targeted system without raising any alerts.
While the Defender has a contrary objective of maximising the Attacker's time spent interacting with the honeypot rather than the real system to gather as much intelligence on the Attacker's activities as possible. 
The Defender's expected utility in HCG is defined by the amount of
\emph{cyber threat intelligence gained} $g(\vulnerabilityselected)$, which depends on the overall time spent by the Attacker exploiting vulnerabilities and the \emph{cost of monitoring} the offered vulnerabilities $c(\vulnerabilityselected)$ i.e.,
\begin{align} \label{eqn:expected-game-utility}
    \util_D = g(\vulnerabilityselected) - c(\vulnerabilityselected) 
\end{align}
Note that, for ease of presentation, $U_D$ is used to express the expected game utility of the Defender regardless of the chosen type of honeypot.

\begin{remark}
    The Defender aims at maximising the overall time spent by the Attacker interacting with the honeypot with the goal to increase cyber threat intelligence. On contrary, the Attacker aims at minimising the time required to exploit a targeted vulnerability.
\end{remark}

%
%

\section{Optimality Analysis}\label{sec:analysis}

This section discusses how \decept determines optimal Defender strategies in the presence of an Attacker that can cause maximum damage as expressed by the zero-sum nature of the Honeypot Configuration Game (HCG).

\subsection{Decision complexity}
We represent HCG in a matrix form with $m\cdot\binom{n}{m}\in O(m\cdot n^m)$ elements which is computationally intractable even for a moderately small $n,m$.
To avoid the combinatorial explosion, we restrict the Defender's choices to only the set $\vulnerabilityset$ of all vulnerabilities from which she can select $n$ vulnerabilities and assign her randomised choices as a probability distribution over $\vulnerabilityset$. 
We represent this in the form of a matrix game with the action space being $\vulnerabilityset$ instead of a family of all $n$-element subsets of $\vulnerabilityset$ transforming the action space to a tractable size $n=\abs{\vulnerabilityset}$.
The equilibrium strategy of the Defender in the mixed extension of this revised game is represented by a vector $\vec x=[p_1, \ldots, p_n]^T \in \Delta(\vulnerabilityset) \subseteq[0,1]^n$, \emph{constrained} to satisfy
\begin{equation}\label{eqn:n-out-of-m-selection}
  \abs{\set{i: p_j\neq 0}}\leq m,
\end{equation}
so that \decept offers the freedom to implement $m$ or less vulnerabilities in the honeypot.

\subsection{Payoffs maximisation}

The zero-sum payoff is derived using the equation \eqref{eqn:expected-game-utility} and depends on the time required to exploit a vulnerability. 
Considering that we do not offer a single vulnerability from $\vulnerabilityselected$ as a pure strategy of the Defender but a set of vulnerabilities $\vulnerabilityset$ to choose from, the game does not satisfy the usual form of a matrix game. 
Considering $t_j$ as the time to exploit vulnerability $v_j$, the expected payoff of the Defender equals
\begin{equation}\label{eqn:naive-adv-rewards}
    u_j := \left\{
            \begin{array}{ll}
            p_j\cdot t_j, & \hbox{if }p_j>0; \\
            \infty, & \hbox{otherwise.}
            \end{array}
        \right.
\end{equation}
This equals either the expected time to exploit a vulnerability times the probability that the vulnerability was offered by the Defender ($p_j>0$) or infinite if the Attacker invests time on creating an exploit for a vulnerability that was not offered by the Defender. 
Since the Attacker aims at minimising the exploitation time $t_j$, any choice leading to $\infty$ would be a dominated strategy hence would never be chosen to be exploited by the Attacker.

In standard game theory with real-valued utility functions, the existence of a Nash equilibrium is assured when the strategy spaces are non-empty compact subsets of a metric space and the utility functions are continuous with respect to the metric \cite{fudenberg_game_1991}. 
It can be observed in \eqref{eqn:naive-adv-rewards} that the expected payoff is an unbounded, even discontinuous function and thus needs to be revised to obtain the equilibrium. 
We address the unbounded-discontinuous nature of the utility function by defining the nature of the Attacker to be a maximiser of the expected residual time obtained for using an exploit. 
This assumption is in line with the rational characteristics of attackers seen in the wild who aim to maximise their payoff from an attack \cite{carroll2011game}.
We choose any constant $T>\max\set{t_j: v_j \in \vulnerabilityset}$, and let the Attacker maximise the expected residual time $T-t_j$, which is by construction equivalent to minimising $t_j$. Substituting $t_j$ by $T-t_j$ in \eqref{eqn:naive-adv-rewards}, we let the Attacker maximise the expected payoff without altering her objective and just changing the optimisation goal. 
The revised expected payoff function of the Attacker is
\begin{equation}\label{eqn:adv-rewards}
    u_j := p_j\cdot (T-t_j),
\end{equation}
whenever the $v_j$-th vulnerability in $\vulnerabilityselected\subseteq \vulnerabilityset$ is chosen. 
Depending on whether vulnerability $v_j$ was offered by the Defender in first place, $u_j$ is either
\begin{itemize}
  \item $u_j>0$, if $p_j>0$, when the vulnerability was offered, or
  \item $u_j=0$, if $p_j=0$, which is a strategically dominated choice for a maximising Attacker.
\end{itemize}
It can be asserted that the revised payoff function \eqref{eqn:adv-rewards} is bounded and no longer has the troublesome discontinuities as observed in \eqref{eqn:naive-adv-rewards}. 
Hereafter, we consider \eqref{eqn:adv-rewards} as the expected payoff of the Attacker and equivalently multiplied by $-1$ as the expected payoff of the Defender.

To highlight the applicability of \decept, we describe and analyse two distinct versions of HCG: (i) \textit{without} the cost of re-configuration (HCG-a); and (ii) \textit{with} the cost of re-configuration (HCG-b). 
HCG-a presents the baseline model of interaction between the Defender and the Attacker with no cost for re-configuring the honeypot in each round of the game.
In HCG-b, we increase the complexity of the model to account for the cost of patching and opening a new vulnerability (re-configuring the honeypot), in each iteration of the game, which affects the overall budget available to the Defender. 
The re-configuration cost moderates the number of rounds the game could be played thus affecting the Defender's expected utility.
Through these games, we investigate the importance of re-configuration cost in the Defender's investment decisions, which was identified by Rass et al. \cite{rass2017cost} as a key component for an acute game-theoretic model. 
A comparative analysis of these games based on the developed use case is discussed in Section \ref{sec:evaluation}.

\subsection{Decision analysis for HCG-a} \label{sec:HCG-a}

This version of HCG presents a baseline model capturing the interaction between the Defender and the Attacker. 
It presents the optimal honeypot configuration when the re-configuration cost is ignored providing insights into the baseline configuration strategies. 
These baseline strategies are then used to assess the impact of the re-configuration cost to the Defender's decisions which are studied by the HCG-b version, presented in the following section.
We let $\vec x=[p_1,\ldots,p_n]\in[0,1]^n$ and $\vec y=[q_1,\ldots,q_n]^T\in[0,1]^n$ both being probability distributions, i.e., constrained to satisfy: (i) $p_k \geq 0, q_k \geq 0, k = {1, 2, \cdots, n},$ and (ii) $\sum_k p_k = 1 ~\text{and}~ \sum_k q_k = 1.$
The payoff of the game depends on the Attacker's probability of choosing a vulnerability to exploit (represented as $q_j$) and the Defender's probability of choosing the targeted vulnerability to offer (represented as $p_i$).
Since both make their choices at random, the expected utility is expressed as
\begin{equation}\label{eqn:expected-utility}
    \util(\vec x, \vec y) = \sum_{i,j} p_i \cdot q_j \cdot \bar{u}_j\stackrel{\eqref{eqn:adv-rewards}}=\sum_{i,j} p_i \cdot q_j\cdot (p_j\cdot [T-t_j]) 
\end{equation}
To be noted from \eqref{eqn:expected-utility} that the expected utility implicitly depends on $i$. The variable $i$ iterates over all choices $i\in \vulnerabilityset=\set{v_1,\ldots,v_n}$ of the Defender, but the payoff to the Attacker occurs if the vulnerability $v_i$ was actually chosen for exploitation.
The expected utility depends on the sum running over $i$ in \eqref{eqn:expected-utility}.

Since the expected payoff continuously depends on the mixed strategies $\vec x, \vec y$ of both players, using Glicksberg's theorem \cite{fudenberg_game_1991}, we have an equilibrium when $\min_{\vec x} \max_{\vec y} ~\util = \max_{\vec y} \min_{\vec x} \util$. 
Towards solving this minimax optimisation, let us rewrite $\util$ in matrix notation.  
To this end, observe that
\begin{align}\label{eqn:dyadic-product}
    \vec x^T\vec x=\left(
                     \begin{array}{cccc}
                       p_1p_1 & p_1p_2 & \cdots & p_1p_n \\
                       p_2p_1 & p_2p_2 & \cdots & p_2p_n \\
                       \vdots & \vdots & \ddots & \vdots \\
                       p_np_1 & p_np_2 & \cdots & p_np_n \\
                     \end{array}
                   \right) \text{ and }& \nonumber \\
                   \left(
      \begin{array}{ccccc}
        q_1(T-t_1) & 0 & 0 & \cdots & 0 \\
        0 & \ddots & 0 & \cdots & 0 \\
        \vdots & \vdots & q_j(T-t_j) & \cdots & 0 \\
        0 & \ddots & \ddots & \ddots & \vdots \\
        0 & \cdots &  & 0 & q_n(T-t_n) \\
      \end{array}
    \right) & \nonumber \\ 
    = \underbrace{\text{diag}[T-t_1,\ldots,T-t_n]}_{=:\payoffmatrix}\cdot\left( \tiny \begin{array}{c}
        q_1 \\
        \vdots \\
        q_n \\
    \end{array}
    \right) = \payoffmatrix\cdot \vec y &
\end{align}
and multiplying $\vec x^T\vec x$ by the diagonal matrix $\payoffmatrix \cdot \vec y$, we technically multiply the $j$-th row in \eqref{eqn:dyadic-product} by the value $q_j(T-t_j)$ for all $j=1,2,\ldots,n$. Now, to reproduce \eqref{eqn:expected-utility}, we are left with adding up the rows in the resulting $n\times 1$ vector, which is doable by a multiplication with the row-vector $\vec e=[1,1,\ldots,1]\in\R^{1\times n}$. 
The matrix representation being $\util(\vec x,\vec y)=\underbrace{\vec e\cdot \vec x\cdot \vec x^T\cdot \payoffmatrix}_{=:\vec A(\vec x)}\cdot\vec y = \vec A(\vec x)\cdot \vec y$, in which $\vec A(\vec x)$ is a $(1\times n)$-matrix for all $\vec x$.

Let the Defender choose $\vec x$. 
The Attacker's problem, following the Defender, is choosing $\vec y$ towards
\begin{equation}\label{eqn:inner-optimization}
    \max_{\vec y} \vec A(\vec x)\cdot\vec y = \max_{1\leq i\leq n} \vec A(\vec x)\vec e_i^{~T},
\end{equation}
since the inner maximisation is the selection of the largest element from the vector $\vec A(\vec x)\in\R^n$, achievable by a discrete optimisation over all unit vectors $\vec e_i$ with a 1-entry only at the $i$-th coordinate and being zero elsewhere.
Introducing the scalar $\nu$ to represent the value of the inner optimisation \eqref{eqn:inner-optimization}, we arrive at the Defender's problem
\begin{align}\label{eqn:optimization-constraints}
    \min\nu \quad \text{ s.t. }
    \begin{cases}
    \nu&\geq \vec A(\vec x)\vec e_i^{~T} \quad\text{for all~}i=1,2,\ldots,n,\\
    \sum_{i=1}^{n}p_i&=1,\\
    p_i&\geq 0.
    \end{cases}
\end{align}
This is a nonlinear problem with smooth objective and constraints. 
The value obtained from this optimisation is used to compute $g(\vulnerabilityselected)$ and $c(\vulnerabilityselected)$, used in equation
\eqref{eqn:expected-game-utility}. 
We define $g(\vulnerabilityselected)$ and $c(\vulnerabilityselected)$ to be logistic functions \cite{hausken2006returns}, 
and are represented as
\begin{align}\label{eqn:logistic-fuction}
    g(\vulnerabilityselected) = e^{ \frac{1}{1+\lambda \cdot \nu_{\vulnerabilityselected}}} \quad \text{ and } \quad c(\vulnerabilityselected) =  e^{\frac{1}{\alpha \cdot \abs{\vulnerabilityselected} \cdot \nu_{\vulnerabilityselected}}}
\end{align}
where $\nu_{\vulnerabilityselected}$ denotes the optimum $\nu$ obtained by solving \eqref{eqn:optimization-constraints} (or \eqref{eqn:reconfiguration-optimisation-constraints} for HCG-b in next section); and $\lambda \in (0,1)$ denotes the learning rate for a honeypot type as detailed in Section \ref{sec:model}.

The $\lambda$ value can also be expressed as the degree of effectiveness of a honeypot which can be represented as \cite{qassrawi2010deception}:
(i) the time required by the Attacker to realise that he is attacking a decoy system; (ii) the type of honeypot implemented to deceive the Attacker; or (iii) the amount of collected attack data. 
As any of the above factors are higher for $\high$, we consider $\lambda$ for $\high$ is larger than $\low$. 
The $\alpha \in (0,1)$ value denotes the average monitoring cost factor of a honeypot type.
$\high$ must be supported with adequate data collection and control mechanisms to ensure reliable adversarial information gathering, and to prevent the honeypot from being used as a foothold to attack connecting devices and networks. 
$\high$ requires significant resources for continuous monitoring and logging all the interactions to determine the Attacker's motives and methods.
Thus, high-interaction honeypots generally exhibit greater monitoring costs than low-interaction honeypots leading to a higher $\alpha$. \\

\ifExamples
%
%
\subsubsection*{Example with HCG-a}\label{section:HCG-a-example}
Let the Defender be the minimising player with $\high$. 
We consider the exploitation time as categorical values similar to the complexity metric of the vulnerabilities in CVSS scores. 
We let the exploitation time for \emph{low, medium}, and \emph{high} complexity vulnerabilities be $1, 2$ and $3$, respectively. 
For example, let $m=3, n=\abs{\vulnerabilityselected}=3$, the associated exploitation time $\text{diag}(\reconmatrix)=\set{2,1,3}$, and $T=4$. 
The payoff matrix $\payoffmatrix$ is constructed as
\begin{equation*}
    \payoffmatrix = \left(
    \begin{array}{ccc}
         T-t_1 & 0 & 0  \\
         0 & T-t_2 & 0  \\
         0 & 0 & T-t_3  \\
    \end{array}
    \right) = \left(
    \begin{array}{ccc}
        2 & 0 & 0 \\
        0 & 3 & 0 \\
        0 & 0 & 1 \\
    \end{array}
    \right)
\end{equation*}
Analysing the game by solving \eqref{eqn:optimization-constraints}, yields $v(\vulnerabilityselected)^* \approx 0.545$ and the result at an equilibrium strategy $x^*_0 \approx (0.273,0.182,0.545)$. 
To compute the expected game utility $\util_D$ using \eqref{eqn:expected-game-utility}, we let the re-configuration cost $\reconcostset = 0$, as we demonstrate this example with HCG-a, $\lambda = 0.6$ and $\alpha = 0.7$. 
Then, using \eqref{eqn:logistic-fuction}, we compute $g(\vulnerabilityselected) \approx 2.124$ and $c(\vulnerabilityselected) \approx 0.418$ leading to $\util_D \approx 1.706$.
\fi

\subsection{Decision analysis for HCG-b}\label{HCG-b-example}
In the following, we extend our analysis to include the re-configuration cost of the Defender in HCG and refer to it as HCG-b. 
Since the Defender aims at making the honeypot attractive to the Attacker, they are required to demonstrate some activities in the honeypot by patching some vulnerabilities and opening up new ones. 
However, such re-configuration incur costs.
For each vulnerability $v_j\in \vulnerabilityselected$, we denote $s_j^+$ as the cost (in time) for ``offering'' vulnerability $v_i$ in the honeypot, and $s_j^-$ as the cost (in time) for patching vulnerability $v_i$, i.e., removing it from the honeypot.
We further consider that if a vulnerability $v_j$ in the current round remains in the honeypot for the next round, no re-configuration cost occurs and $c^+_i$ is zero.
Likewise, if the vulnerability $v_j$ is not offered in this round of the repeated game nor included in the next round, then $c^-_j$ is zero. 
Assuming stochastic independence of the rounds in the game, we investigate the cases where:
\begin{enumerate}
    \item[]\textit{Case 1}. vulnerability $v_j$ was offered in the current round with probability $p_j$ and is patched in the next round, with probability $1-p_j$. Thus, the expected re-configuration cost is $p_j\cdot (1-p_j)\cdot s_j^-$. \\
    \item[]\textit{Case 2}. vulnerability $v_j$ was not offered in the current round with probability $1-p_j$ and the honeypot will be configured to have $v_j$ in the next round leading to the expected re-configuration cost of $(1-p_j)\cdot p_i\cdot s_j^+$.
\end{enumerate}
Since the coefficients of the expected re-configuration cost in both cases above are the same, we end up with the re-configuration penalty term to be the sum of all $p_j\cdot (1-p_j)\cdot (s_j^++s_j^-)$ over $v_j\in \vulnerabilityselected$. 
The matrix notation of the re-configuration cost $\reconcostset(\vulnerabilityselected)$ can be expressed as
\begin{align}\label{eqn:reconfiguration-cost-function}
&\vec x^T\cdot \reconmatrix \cdot(\vec 1-\vec x), \nonumber \\
\text{ where } & \reconmatrix =\begin{pmatrix}
    s_1^++s_1^- & 0 & \dots & 0 \\
    0 & s_2^++s_2^- & \dots & 0 \\
    \vdots & \vdots & \ddots & \vdots \\
    0 & 0 & \dots & s_n^++s_n^-
  \end{pmatrix} 
\end{align}
in which $\vec 1$ is the vector of all ones, and $\reconmatrix$ is a diagonal matrix with the re-configuration cost.

The re-configuration cost matrix $\reconmatrix$ can be included in the optimisation, as detailed in \cite{rass2017cost}, as a penalty term $\vec x^T \cdot \reconmatrix \cdot (\vec 1-\vec x)$ to the expected payoff function.
To express the trade-off between the payoff matrix $\vec A(\vec x)$ and the re-configuration cost matrix $\reconmatrix$, we introduce a smoothing factor $0<\beta<1$.
The so-enhanced optimisation problem can be expressed as: 
\begin{align}\label{eqn:reconfiguration-optimisation-constraints}
    & \min\nu \nonumber \\
    \text{such that}& \nonumber \\
    &\nu\geq \beta\cdot A(\vec x)\vec e_i + (1-\beta) \cdot \vec x^T \cdot \reconmatrix \cdot (\vec 1-\vec x) \nonumber \\
    & \qquad \forall i=1,2,\cdots,n, \nonumber \\
    & \vec 1^T\vec x =1, \nonumber \\
    & \vec x \geq 0.
\end{align}

\decept can also incorporate a more complex HCG where all $m$-out-of-$n$ subsets are included as separate strategies to re-configure the honeypot. 
For example, let the set of vulnerabilities in the current round be $\vulnerabilityselected_i$ and the set of vulnerabilities for the next round be $\vulnerabilityselected_j$. 
Then the cost to switch from $i$ to $j$, i.e., re-configuration cost, is $s_{ij} = \sum_{k\in \vulnerabilityselected_j\setminus \vulnerabilityselected_i} s_k^+ + \sum_{k\in \vulnerabilityselected_i\setminus \vulnerabilityselected_j} s_k^-$, with $i,j$ ranging over all strategies, forming a matrix $\reconmatrix=(s_{ij})$.
According to \cite{rass2017cost}, the penalty term to go into the optimisation is then the plain quadratic form $\vec x^T \cdot \reconmatrix \cdot (1-\vec x)$. \\

\ifExamples
%
%
\subsubsection*{Example with HCB-b}
We extend the example in Section \ref{section:HCG-a-example} by considering the re-configuration cost. 
We adopt the concept of time that uses categorical values similar to the complexity metric of the vulnerabilities in the CVSS score.
We let the re-configuration cost for \emph{low, medium} and \emph{high} complexities to $1,2 \text{ and } 3$, respectively. 
Let the Defender be the minimising player with $\high$ and let $m=3, n=\abs{\vulnerabilityselected}=3$, the associated exploitation time $\text{diag}(\reconmatrix)=\set{2,1,3}$, and $T=4$. Then, the payoff and the re-configuration cost matrices are given by
\begin{equation*}
    \payoffmatrix = \left(
    \begin{array}{ccc}
        2 & 0 & 0 \\
        0 & 3 & 0 \\
        0 & 0 & 1 \\
    \end{array}
    \right) ,
    \quad \text{ and }\quad
    \reconmatrix = \left(
    \begin{array}{ccc}
        2 & 0 & 0 \\
        0 & 1 & 0 \\
        0 & 0 & 3 \\
    \end{array}
    \right)
\end{equation*}
Analysing the game in the described way and solving \eqref{eqn:reconfiguration-optimisation-constraints}, yields $v(\vulnerabilityselected)^* \approx 0.562$ at an equilibrium strategy $\vec x^*_0 \approx (0.332,0.304,0.364)$. 
Naturally, the Defender's expected loss $f_v$ here is higher than the conventional game without re-configuration cost (see \ref{section:HCG-a-example}).
To compute the expected game utility $\util_D$ using \eqref{eqn:expected-game-utility}, we let $\lambda = 0.6$, $\alpha = 0.7$ and $\beta = 0.5$ leading to $g(\vulnerabilityselected) \approx 2.112$ and $c(\vulnerabilityselected) \approx 0.429$. 
The re-configuration cost for offering these vulnerabilities can be obtained using $(\vec x^*_0)^T\cdot \reconmatrix \cdot (1-\vec x^*_0) \approx 1.35$. 
Then, taking the re-configuration cost into account gives $\util_D \approx 0.33$. 

\begin{remark}
    When compared to the example in Section \ref{section:HCG-a-example}, it is evident that the re-configuration cost is important in determining the number of playable rounds of the HCG and eliminating the unattainable equilibrium.
\end{remark}
\fi

\subsection{Scalability analysis}

To evaluate the scalability of \decept, we ran simulations on a 2.8GHz Intel core i7 with 16GB RAM using Python 3.7.0 with the \emph{scipy.optimize} package\footnote{\url{https://docs.scipy.org/doc/scipy/reference/tutorial/optimize.html}}.
Our experiments showed that \decept can solve well up to and at least $n=1000$ vulnerabilities, but has the issue with the optimum assigning of positive probabilities $p_j>0$ for all 1000 vulnerabilities with the time of exploits ($t_j$) being represented by a random sampled set of values from $\set{1,2,3}$.

In a real-world situation, configuring the honeypot with all vulnerabilities in $\vulnerabilityset$ will be overly laborious with significant maintenance and monitoring costs.
This was evident from the fact that even choosing vulnerabilities with $p_j < 0.001$, we ended up with 319 vulnerabilities in one of the experiments. 
Offering such a large set of vulnerabilities in a honeypot might be infeasible, let alone economic.
Our goal was to keep a small subset of vulnerabilities open, to avoid exposing the honeypot and to unnecessary suspicion from the Attacker.
Acknowledging the possibility of the honeypot to be identified by the Attacker, we introduced the constraint \eqref{eqn:n-out-of-m-selection} to enforce an $m$-out-of-$n$ selection from $\vulnerabilityset$. 
This reduces the problem into one with cardinality constraints, to which specialised approximation methods are applicable \cite{ruiz2010optimization}. 
Our experiments with simple methods to replace \eqref{eqn:n-out-of-m-selection}, such as entropy constraints or smooth approximation for the indicator function, has failed when a large vulnerability set (around $n=100$) is considered unless $m$ is as large as $n$.
However, if the honeypot is used exclusively to monitor unauthorised access, the constrain on the number of configured vulnerabilities could be exempted. 

The expected game utility is achieved by the following steps:
\begin{enumerate}
  \item Solve the optimisation problem as stated above (with or without the cardinality constraint depending on computational feasibility), and call the output value $\vec x=(p_1,\ldots,p_n)$.
  \item Iterate over all $v_j\in \vulnerabilityselected \subseteq \vulnerabilityset$, and with probability $p_j$, as determined from the optimisation problem, choose to include a vulnerability in the honeypot.
  \item Evaluate $\vec A(\vec x)$ and determine the Attacker's optimal (Defender's worst-case) strategy as the maximum over the elements of $\vec A(\vec x)\in\R^{n}$.
  \item Compute $\nu$ and the re-configuration cost $\vec x^{*T} \cdot \reconmatrix \cdot (1-\vec x^*)$ at the optimum $\vec x^*$ to calculate $\util_D$ in \eqref{eqn:expected-game-utility}.
\end{enumerate}

%
%

\begin{table*}[ht]
    \centering
    \resizebox{\textwidth}{!}{\begin{tabular}{|c|c|c|c|c|c|c|c|}
        \hline
        $\#$ & $v_1$ & $v_2$ & $v_3$ & $v_4$ & $v_5$ & $v_6$ & $v_7$ \tabularnewline
        \hline 
        \textbf{CVE ID} & CVE-2018-9311 & CVE-2018-9318 & CVE-2019-9977 &  CVE-2018-9313 & CVE-2012-6510 & CVE-2018-6508 & CVE-2016-9337 \tabularnewline
        \hline
        \textbf{Score} & 10 & 10 & 6.8 & 5.7 & 4.3 & 6 & 4  \tabularnewline
        \hline
        \textbf{Access} & Remote & Remote & Remote & Remote & Remote & Remote & Remote \tabularnewline
        \hline
        \textbf{Complexity} & Low & Low & Medium & Medium & Medium & Medium & High \tabularnewline
        \hline
    \end{tabular}}
    \caption[]{Sample set of vehicular security vulnerabilities with \ac{CVSS} metrics.}
    \label{tab:cve-snapshot}
\end{table*}

\section{Case study}\label{sec:evaluation}

This section presents a case study where \decept is assessed using known vulnerabilities related to autonomous and connected vehicles. 
For the purpose of this case study, we consider the vulnerabilities from the Common Vulnerabilities and Exposure (CVE) list found within the National Institute of Standards and Technology (NIST) \ac{NVD}\footnote{\url{https://nvd.nist.gov/vuln/data-feeds}}.
The CVE data includes a description, Common Vulnerability Scoring System (CVSS) base scores, vulnerable product configuration, and weaknesses' categorisation information on each identified vulnerability.
We primarily utilise the CVSS metrics to acquire parametric values required for \decept.
CVSS is a publicly available industry standard that details the characteristics and severity of software vulnerabilities and is built upon three core metric groups: Base, Temporal, and Environment. 
The \textit{Base metric} represents the intrinsic qualities of a vulnerability that remain unchanged over time and across user environments. 
The \textit{Temporal metric} reflects the characteristics of a vulnerability that can change over time, while the \textit{Environmental metric} reflects qualities of a vulnerability that are unique to a user's environment.
This case study uses the Base metrics to extract the parameters to be used in \decept. 

First, we consider a small sample of vulnerabilities to assess \decept.
Nevertheless, \decept can be implemented with any number of vulnerabilities to offer decision support to the Defender. 
Next, the \ac{NVD} is checked for available patches for a vulnerability.
Identifying vulnerabilities with available patches is critical to \decept as without patches it is infeasible to re-configure the honeypot. 
Once a vulnerability with patch has been identified, we obtain its \ac{CVSS} metrics. 
Finally, \decept is applied to obtain the optimal honeypot configuration for the versions of the Honeypot Configuration Game (HCG) presented in Section \ref{sec:analysis}. 

\subsection{Data and use case composition}

Taking advantage of the complexity metric of the \ac{CVSS}, we derive the exploitation time $t_j$ and the re-configuration cost $\reconcostset(v_j)$ for a vulnerability $v_j$. 
The complexity metric expresses the \emph{anticipated efforts} needed to exploit a vulnerability. 
We associate a \emph{low} complexity to ``short'' time, a \emph{medium} complexity to ``medium'' time and a \emph{high} complexity to ``long'' time. 
With such association, we set $t_j, s^+_j ,s^-_j \in \set{1,2,3}$ for a vulnerability $v_j$ within the ranks of the categorical values of the complexity metric. 
We further use the complexity metric to distinguish vulnerabilities that can be supported by \ac{LIH} and \ac{HIH}. 
Vulnerabilities requiring higher effort to exploit require higher access privileges which can only be supported through high-interaction honeypots. 
We, thus, consider that a low-interaction honeypot can only support vulnerabilities with \emph{low} and \emph{medium} complexities. 
Table \ref{tab:cve-snapshot} presents the list of security vulnerabilities used from the \ac{CVE} database and relevant \ac{CVSS} metrics. 
The \ac{CVSS} score ranges from 0 to 10 and specifies the potential impact of a vulnerability. 
For example, a vulnerability providing access to the breaking system of a vehicle will have higher impact compared to the one that compromises the windscreen wipers.  


We assume that the Defender can offer at most $m = \abs{\vulnerabilityselected} = 6$ vulnerabilities through a honeypot.
The Honeypot Configuration Game (HCG) is played for a finite number of rounds equivalent to the maximum number of vulnerabilities to be offered by a honeypot.  
Naturally, honeypot monitoring cost increases with the offered number of vulnerabilities.
Deploying \decept assists the Defender in determining the right type of honeypot to implement and the optimum number of vulnerabilities to offer optimising the expected utility.
To compute the expected game utility ($U_D$), we first set values for the time constant $T$, exploitation times $t_j$ and re-configuration cost $\reconcostset(v_j)$ for all vulnerabilities $v_j \in \vulnerabilityset$.

\begin{table}[ht]
    \centering
    \resizebox{\columnwidth}{!}{\begin{tabular}{|c|c|c|c|c|c|c|c|}
        \hline
        $v_j$ & $v_1$ & $v_2$ & $v_3$ & $v_4$ & $v_5$ & $v_6$ & $v_7$ \tabularnewline
        \hline 
        $t_j$ & 1 & 1 & 2 & 2 & 2 & 2 & 3  \tabularnewline
        \hline
        $T-t_j$ & 3 & 3 & 2 & 2 & 2 & 2 & 1  \tabularnewline
        \hline
        $\reconcostset(v_j)$ & 1 & 1 & 2 & 2 & 2 & 2 & 3 \tabularnewline
        \hline
    \end{tabular}}
    \caption{Exploitation time and re-configuration cost for sample vulnerabilities.}
    \label{tab:usecase-table}
\end{table}

As detailed earlier, we heuristically consider the ``complexity'' metric to derive these values.
We let $T:=4$ as any choice for $T>\max\set{1,2,3}$ is admissible. 
Table \ref{tab:usecase-table} presents the $t_j$ and $\reconcostset(v_j)$ for the vulnerabilities listed in Table \ref{tab:cve-snapshot}. 
We further consider that the re-configuration cost of a vulnerability equals its exploitation time. 
From the vulnerabilities sample in Table \ref{tab:cve-snapshot}, vulnerability $v_7$ being of high complexity cannot be offered through $\low$ implying that $\vulnerabilityselected = \vulnerabilityset \setminus \set{v_7}$ for $\low$. 
For proportionality, we consider $\vulnerabilityselected = \vulnerabilityset \setminus \set{v_1}$ for $\high$. 
Using these vulnerability sets, we then construct the respective $\payoffmatrix$ and $\reconmatrix$ required to complete the optimisation as introduced in Section \ref{sec:analysis}.

\subsection{Results}

The $\nu$ value, obtained from the optimisation, is a key parameter in determining the honeypot configuration strategy. 
It expresses the maximum of the minimum residual time ($T-t_j$) that the Defender can achieve regardless of the vulnerability chosen by the Attacker to exploit.
The $\nu$ value is influenced by (i) the total number of available vulnerabilities; (ii) the number of vulnerabilities selected to offer; and (iii) the type of the offered vulnerabilities. 
The $\nu$ values decrease with the increase in the number of offered vulnerabilities as with increased number of offered vulnerabilities the Defender has greater chances to deceive the Attacker. 

\begin{figure*}[t]
    \centering
    \begin{subfigure}[b]{0.3\textwidth}
        \includegraphics[width=\textwidth]{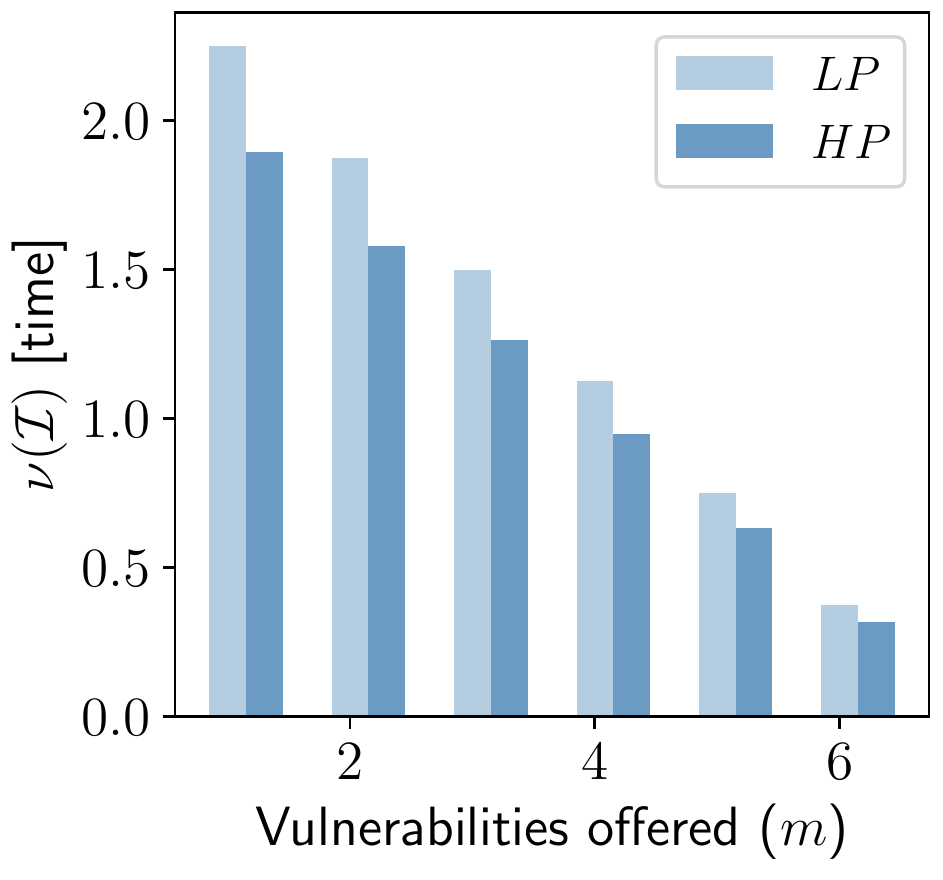}
        \caption{$\nu(\vulnerabilityselected)$ with $\low$ and $\high$}
        \label{fig:m1_usecase_plot1_nu}
    \end{subfigure}
    \begin{subfigure}[b]{0.3\textwidth}
        \includegraphics[width=\textwidth]{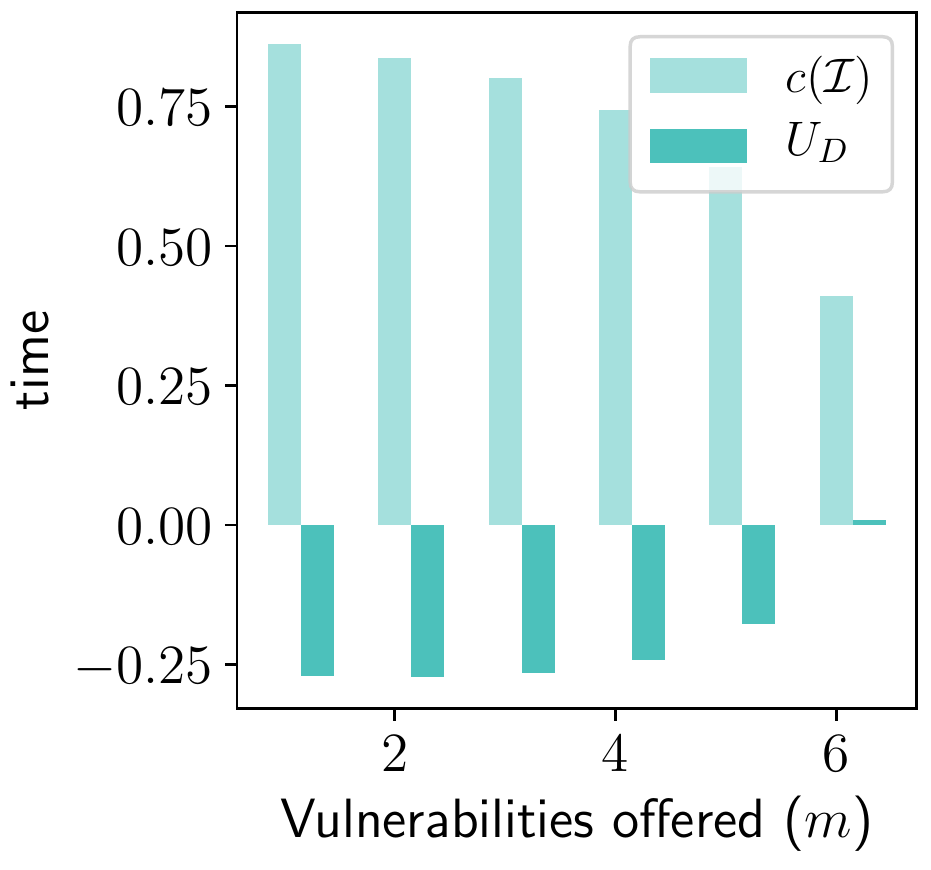}
        \caption{$c(\vulnerabilityselected)$ and $U_D$ with $\low$}
        \label{fig:m1_usecase_plot1_L}
    \end{subfigure}
    \begin{subfigure}[b]{0.3\textwidth}
        \includegraphics[width=\textwidth]{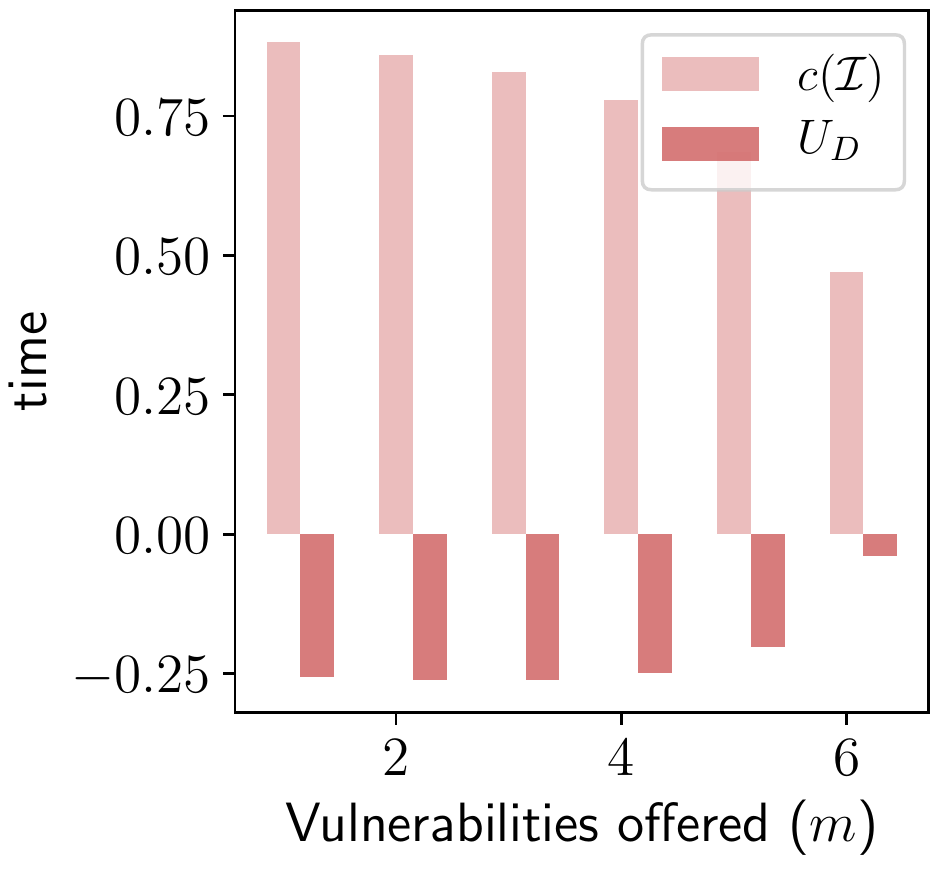}
        \caption{$c(\vulnerabilityselected)$ and $U_D$ with $\high$}
        \label{fig:m1_usecase_plot1_H}
    \end{subfigure}
    \caption{Comparison of HCG-a game utility $\util_D$ and $c(\vulnerabilityselected)$ for a total of six vulnerabilities with $\beta = 0.5$, $\lambda$ being $0.4$ and $0.6$, $\alpha$ being $0.5$ and $0.7$ for $\low$ and $\high$, respectively.}
    \label{fig:m1_usecase_plot1}
\end{figure*}

Figure \ref{fig:m1_usecase_plot1_nu} presents the $\nu$ value over the range $0<m\leq n$ with $\low$ and $\high$ without the re-configuration cost i.e., for HCG-a. 
The minimal value for $\nu$ is attained at $m=n=6$.
Further, Figure \ref{fig:m1_usecase_plot1_L} shows the expected game utility $U_D$ and the honeypot monitoring cost $c(\vulnerabilityselected)$ for HCG-a with $\low$ over the range $0<m\leq n$. 
As expected, the cost of monitoring the honeypot and the gather adversarial activities increases with $\nu$.
In particular, $\nu$ increases with an increase in the number of rounds of the game which enforces the cost of monitoring the honeypot over the duration of the game. 
On the contrary to the escalating monitoring cost, a higher $\nu$ is preferred as the cyber cyber threat intelligence of the Defender grows with the number of rounds of the game. Thus, an optimal choice would be to select $m$ such that $U_D$ is positive and possibly the largest.
The only positive $U_D$ is at $m=6$ suggesting that the expected game utility is the best when offering all six vulnerabilities at once with $\low$. 
With this strategy, the minimum improvement in $U_D$ is approximately $100\%$ compared to any others selection of $m$. Similarly, it can be inferred from Figure \ref{fig:m1_usecase_plot1_H}, which presents $U_D$ and $c(\vulnerabilityselected)$ with $\high$, that implementing $\high$ will be expensive as $U_D$ is always negative regardless of the size of $m$.

\begin{result}
    In \decept, when trying to maximise the duration of engagement with the attacker, an ideal choice for the Defender would be to offer all available vulnerabilities in one round of the game. 
    This strategy particularly supports the objective of wasting attackers' time before throwing them out of the network and confirming the reasoning behind the use of honeypots to dissuade attackers from critical infrastructure and vehicles in IoV. 
\end{result}

\begin{figure*}[t]
    \centering
    \begin{subfigure}[b]{0.3\textwidth}
        \includegraphics[width=\textwidth]{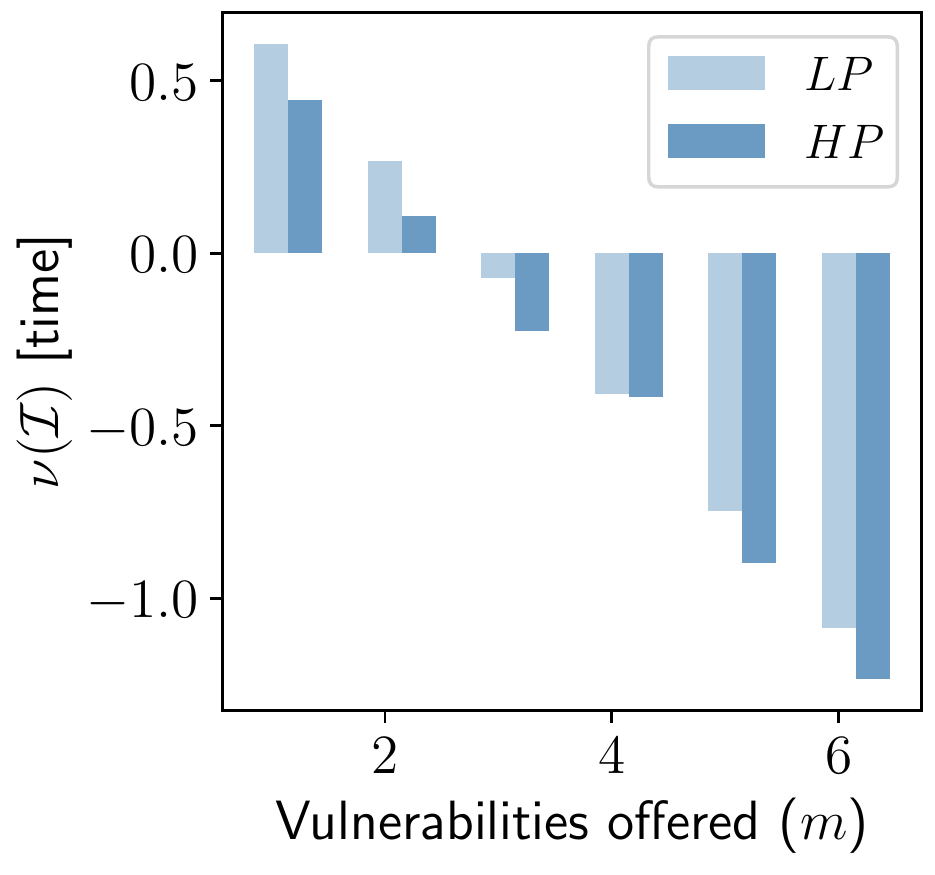}
        \caption{$\nu(\vulnerabilityselected)$ with $\low$ and $\high$}
        \label{fig:m2_usecase_plot1_nu}
    \end{subfigure}
    \begin{subfigure}[b]{0.3\textwidth}
        \includegraphics[width=\textwidth]{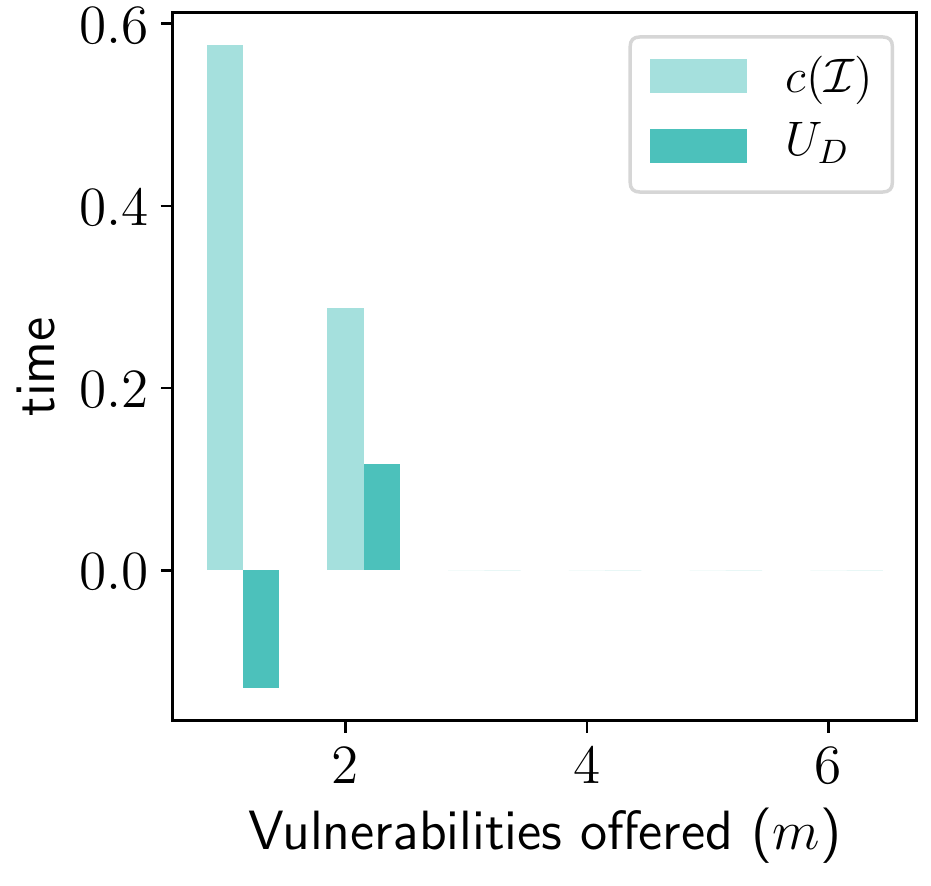}
        \caption{$c(\vulnerabilityselected)$ and $U_D$ with $\low$}
        \label{fig:m2_usecase_plot1_L}
    \end{subfigure}
    \begin{subfigure}[b]{0.3\textwidth}
        \includegraphics[width=\textwidth]{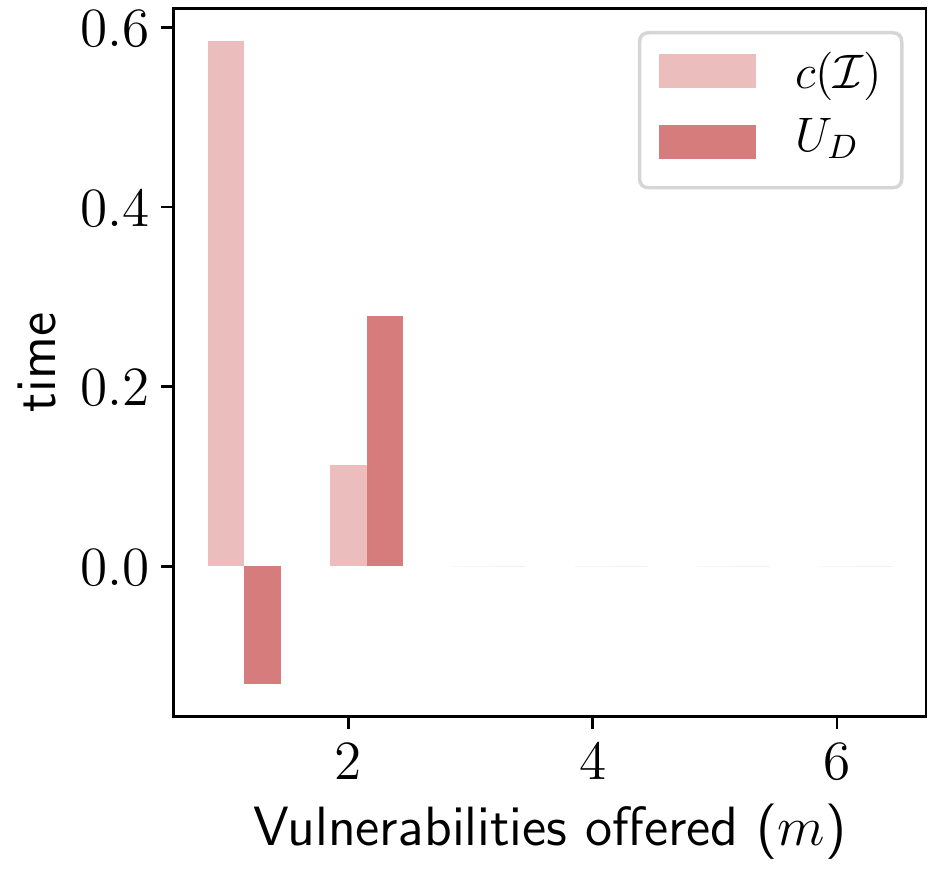}
        \caption{$c(\vulnerabilityselected)$ and $U_D$ with $\high$}
        \label{fig:m2_usecase_plot1_H}
    \end{subfigure}
    \caption{Comparison of the HCG-b game utility $\util_D$ and $c(\vulnerabilityselected)$ for a total of six vulnerabilities with $\beta = 0.5$, $\lambda$ being $0.4$ and $0.6$, $\alpha$ being $0.5$ and $0.7$ for $\low$ and $\high$, respectively.}
    \label{fig:m2_usecase_plot1}
\end{figure*}

The results of the case study with HCG-b are presented in Figure \ref{fig:m2_usecase_plot1}. 
With regards to re-configuration cost, the Defender has to endure additional cost
for every selection of $m$ vulnerabilities.
Figure \ref{fig:m2_usecase_plot1_nu} shows that $\nu$ is negative when $m>2$ implying that the Defender should not offer more than two vulnerabilities in a round. 
The reason for this stems from the fact that HCG is a zero-sum game and $\nu<0$ implies a better payoff for the Attacker. 
Figures \ref{fig:m2_usecase_plot1_L} and \ref{fig:m2_usecase_plot1_H} illustrate the $U_D$ and $c(\vulnerabilityselected)$ with $\low$ and $\high$, respectively. 
According to the results, the Defender gains a better $U_D$ with $\high$ when $m=2$. 
The Defender, while spending approximately $62\%$ less on monitoring cost, can achieve an improved $U_D$ of approximately $134\%$ compared to the best strategy with $\low$ which also happens to be with $m=2$. 
\begin{result}
    When trying to maximise the cyber threat intelligence, the honeypot configuration should be such that it engages the Attacker longer leading to a positive expected game utility. 
    \decept assists in determining the optimal number of vulnerabilities to be offered to gain the largest positive expected utility. 
    The positive expected utility reflects that the collected amount of cyber threat intelligence is higher than the cost of monitoring the honeypot leading to a positive Return on Security Investment (ROSI).
\end{result}

\begin{result}
    The re-configuration cost can be seen as a cost for switching from one Nash Equilibrium of the game to another. It refines the duration of the game by eliminating unplayable choices regarding the number of vulnerabilities to be offered in a round. 
    \decept, through such strategies, recommends optimal honeypot deception investment plans to engage attackers and achieve cyber threat intelligence.
\end{result}

%
%

\section{Conclusions}\label{sec:conclusion}

The application of honeypots is a promising approach for protecting IoV networks. 
If an attacker is successfully lured by the honeypot, the adversarial activities captured by the honeypot can be used to learn about the attacker's motives and techniques.
Successively, this knowledge, also referred to as cyber threat intelligence, can contribute to protecting existing system components by improving intrusion detection with new attack signatures or anomalous behaviour deviating from norms of protocols and systems' behaviours.
Besides the inevitable cost of maintaining IoV honeypots, it is a challenge to design convincing honeypots to successfully deceive attacks
This paper proposed a novel framework called \decept which aims to assist IoV network administrators with the optimal configurations and investments in honeypots.
\decept is built upon two models: (i) a formal model of assessing the option of the Defender to invest in cyber deception using honeypots; and (ii) a game-theoretic model to strategically determine the configuration and selection of honeypot to be deployed in IoV network.
\decept empowers the network administrator to derive optimal decisions regarding honeypot deception based on an available budget. 
We take into consideration the number and type of vulnerabilities to be offered by the honeypot, the benefit and cost of implementing a vulnerability, the cost for re-configuring a honeypot and the available budget for investment in deception.

We demonstrate and evaluate \decept using autonomous and connected vehicular security vulnerabilities collected from the Common Vulnerabilities and Exposure (CVE) data found within the National Vulnerability Database (NVD) and the respective Common Vulnerability Scoring System (CVSS) metrics.
Our evaluation suggests that \decept is capable of supporting IoV network administrators to determine the optimal configuration of honeypots for (i) wasting attacker's time and (ii) maximising the collection of cyber threat intelligence.
\decept also highlighted the importance of re-configuration cost in determining the duration of the game by eliminating unplayable choices leading to a realistic investment plan. 


A key question future work could investigate is when to re-configure the honeypot. 
An option would be to consider finite horizon games where the vulnerability set becomes exhausted at some point leading to reopening a vulnerability that has been closed in the past. However, such an action could rise suspicion and the deception might fail. The Defender's option could be to invest more to introduce entirely new set of vulnerabilities or refresh the honeypot.
Formally, the game can be expressed in extensive form, with a number of stages that correspond to the number of re-configurations allowed given the vulnerability set. 
This will introduce more solution concepts like Subgame Perfect Nash equilibria or even Stackelberg Nash equilibria. 
Last, future work can include approaches that combine both game theory and machine learning to develop defensive deception techniques. In such a hybrid approach, reinforcement learning with game theory could be used to formulate players' utility functions, estimate opponent's beliefs and update the optimal strategy and predict opponent's actions by analysing data from host vehicles, network and threat actor behaviours.



\ifCLASSOPTIONcaptionsoff
  \newpage
\fi

\begin{acronym}
\acro{CVE}{Common Vulnerabilities and Exposures}%
\acro{CVSS}{Common Vulnerability Scoring System}%
\acro{NVD}{National Vulnerability Database}%
\acro{LIH}{\textit{low-interaction} honeypot}%
\acro{HIH}{\textit{high-interaction} honeypot}%
\end{acronym}



\bibliographystyle{IEEEtran}
\bibliography{references}

\begin{thebibliography}{10}
\providecommand{\url}[1]{#1}
\csname url@samestyle\endcsname
\providecommand{\newblock}{\relax}
\providecommand{\bibinfo}[2]{#2}
\providecommand{\BIBentrySTDinterwordspacing}{\spaceskip=0pt\relax}
\providecommand{\BIBentryALTinterwordstretchfactor}{4}
\providecommand{\BIBentryALTinterwordspacing}{\spaceskip=\fontdimen2\font plus
\BIBentryALTinterwordstretchfactor\fontdimen3\font minus
  \fontdimen4\font\relax}
\providecommand{\BIBforeignlanguage}[2]{{%
\expandafter\ifx\csname l@#1\endcsname\relax
\typeout{** WARNING: IEEEtran.bst: No hyphenation pattern has been}%
\typeout{** loaded for the language `#1'. Using the pattern for}%
\typeout{** the default language instead.}%
\else
\language=\csname l@#1\endcsname
\fi
#2}}
\providecommand{\BIBdecl}{\relax}
\BIBdecl

\bibitem{kaiwartya2016internet}
O.~Kaiwartya, A.~H. Abdullah, Y.~Cao, A.~Altameem, M.~Prasad, C.-T. Lin, and
  X.~Liu, ``Internet of vehicles: Motivation, layered architecture, network
  model, challenges, and future aspects,'' \emph{IEEE Access}, vol.~4, pp.
  5356--5373, 2016.

\bibitem{gerla2014internet}
M.~Gerla, E.-K. Lee, G.~Pau, and U.~Lee, ``Internet of vehicles: From
  intelligent grid to autonomous cars and vehicular clouds,'' in \emph{2014
  IEEE world forum on internet of things (WF-IoT)}.\hskip 1em plus 0.5em minus
  0.4em\relax IEEE, 2014, pp. 241--246.

\bibitem{CPA15}
G.~Loukas, \emph{Cyber-physical attacks: A growing invisible threat}.\hskip 1em
  plus 0.5em minus 0.4em\relax Butterworth-Heinemann, 2015.

\bibitem{cheng2015routing}
J.~Cheng, J.~Cheng, M.~Zhou, F.~Liu, S.~Gao, and C.~Liu, ``Routing in internet
  of vehicles: A review,'' \emph{IEEE Transactions on Intelligent
  Transportation Systems}, vol.~16, no.~5, pp. 2339--2352, 2015.

\bibitem{patel2017honeypot}
P.~Patel and R.~Jhaveri, ``A honeypot scheme to detect selfish vehicles in
  vehicular ad-hoc network,'' in \emph{Computing and Network
  Sustainability}.\hskip 1em plus 0.5em minus 0.4em\relax Springer, 2017, pp.
  389--401.

\bibitem{verendel2008approach}
V.~Verendel, D.~K. Nilsson, U.~E. Larson, and E.~Jonsson, ``An approach to
  using honeypots in in-vehicle networks,'' in \emph{2008 IEEE 68th Vehicular
  Technology Conference}.\hskip 1em plus 0.5em minus 0.4em\relax IEEE, 2008,
  pp. 1--5.

\bibitem{gantsou2014toward}
D.~Gantsou and P.~Sondi, ``Toward a honeypot solution for proactive security in
  vehicular ad hoc networks,'' in \emph{Future Information Technology}.\hskip
  1em plus 0.5em minus 0.4em\relax Springer, 2014, pp. 145--150.

\bibitem{schmitzstrategy}
Y.~M. Schmitz, ``A strategy for vehicular honeypots.''

\bibitem{la2016deceptive}
Q.~D. La, T.~Q. Quek, J.~Lee, S.~Jin, and H.~Zhu, ``Deceptive attack and
  defense game in honeypot-enabled networks for the internet of things,''
  \emph{IEEE Internet of Things}, vol.~3, no.~6, pp. 1025--1035, 2016.

\bibitem{zhu2021survey}
M.~Zhu, A.~H. Anwar, Z.~Wan, J.-H. Cho, C.~Kamhoua, and M.~P. Singh, ``A survey
  of defensive deception: Approaches using game theory and machine learning,''
  \emph{IEEE Communications Surveys \& Tutorials}, 2021.

\bibitem{la2016game}
Q.~D. La, T.~Q. Quek, and J.~Lee, ``A game theoretic model for enabling
  honeypots in iot networks,'' in \emph{2016 IEEE International Conference on
  Communications (ICC)}.\hskip 1em plus 0.5em minus 0.4em\relax IEEE, 2016, pp.
  1--6.

\bibitem{manshaei2013game}
M.~H. Manshaei, Q.~Zhu, T.~Alpcan, T.~Bac{\c{s}}ar, and J.-P. Hubaux, ``Game
  theory meets network security and privacy,'' \emph{ACM Computing Surveys
  (CSUR)}, vol.~45, no.~3, pp. 1--39, 2013.

\bibitem{do2017game}
C.~T. Do, N.~H. Tran, C.~Hong, C.~A. Kamhoua, K.~A. Kwiat, E.~Blasch, S.~Ren,
  N.~Pissinou, and S.~S. Iyengar, ``Game theory for cyber security and
  privacy,'' \emph{ACM Computing Surveys}, vol.~50, no.~2, pp. 1--37, 2017.

\bibitem{panaousis2017game}
E.~Panaousis, E.~Karapistoli, H.~Elsemary, T.~Alpcan, M.~Khuzani, and A.~A.
  Economides, ``Game theoretic path selection to support security in
  device-to-device communications,'' \emph{Ad Hoc Networks}, vol.~56, pp.
  28--42, 2017.

\bibitem{pibil2012game}
R.~P{\'\i}bil, V.~Lis{\`y}, C.~Kiekintveld, B.~Bo{\v{s}}ansk{\`y}, and
  M.~P{\v{e}}chou{\v{c}}ek, ``Game theoretic model of strategic honeypot
  selection in computer networks,'' in \emph{International Conference on
  Decision and Game Theory for Security}.\hskip 1em plus 0.5em minus
  0.4em\relax Springer, 2012, pp. 201--220.

\bibitem{sakiz2017survey}
F.~Sakiz and S.~Sen, ``A survey of attacks and detection mechanisms on
  intelligent transportation systems: {VANET}s and iov,'' \emph{Ad Hoc
  Networks}, vol.~61, pp. 33--50, 2017.

\bibitem{ring2015connected}
T.~Ring, ``Connected cars--the next target for hackers,'' \emph{Network
  Security}, vol. 2015, no.~11, pp. 11--16, 2015.

\bibitem{lam2021ant}
K.-Y. Lam, S.~Mitra, F.~Gondesen, and X.~Yi, ``Ant-centric iot security
  reference architecture--security-by-design for satellite-enabled smart
  cities,'' \emph{IEEE Internet of Things Journal}, 2021.

\bibitem{aloqaily2019intrusion}
M.~Aloqaily, S.~Otoum, I.~Al~Ridhawi, and Y.~Jararweh, ``An intrusion detection
  system for connected vehicles in smart cities,'' \emph{Ad Hoc Networks},
  vol.~90, p. 101842, 2019.

\bibitem{rivas2011security}
D.~A. Rivas, J.~M. Barcel{\'o}-Ordinas, M.~G. Zapata, and J.~D. Morillo-Pozo,
  ``Security on {VANET}s: Privacy, misbehaving nodes, false information and
  secure data aggregation,'' \emph{Journal of Network and Computer
  Applications}, vol.~34, no.~6, pp. 1942--1955, 2011.

\bibitem{emara2015evaluation}
K.~Emara, W.~Woerndl, and J.~Schlichter, ``On evaluation of location privacy
  preserving schemes for {VANET} safety applications,'' \emph{Computer
  Communications}, vol.~63, pp. 11--23, 2015.

\bibitem{wazid2019akm}
M.~Wazid, P.~Bagga, A.~K. Das, S.~Shetty, J.~J. Rodrigues, and Y.~Park,
  ``Akm-iov: Authenticated key management protocol in fog computing-based
  internet of vehicles deployment,'' \emph{IEEE Internet of Things Journal},
  vol.~6, no.~5, pp. 8804--8817, 2019.

\bibitem{sharma2018survey}
S.~Sharma and A.~Kaul, ``A survey on intrusion detection systems and honeypot
  based proactive security mechanisms in {VANET}s and {VANET} cloud,''
  \emph{Vehicular communications}, vol.~12, pp. 138--164, 2018.

\bibitem{barrett2018framework}
M.~Barrett, ``\BIBforeignlanguage{en}{Framework for improving critical
  infrastructure cybersecurity version 1.1},'' 2018-04-16 2018.

\bibitem{rowe2006measuring}
N.~C. Rowe, ``Measuring the effectiveness of honeypot
  counter-counterdeception,'' in \emph{Proceedings of the 39th Annual Hawaii
  International Conference on System Sciences}, vol.~6.\hskip 1em plus 0.5em
  minus 0.4em\relax IEEE, 2006, pp. 129c--129c.

\bibitem{fielder2016decision}
A.~Fielder, E.~Panaousis, P.~Malacaria, C.~Hankin, and F.~Smeraldi, ``Decision
  support approaches for cyber security investment,'' \emph{Decision Support
  Systems}, vol.~86, pp. 13--23, 2016.

\bibitem{nespoli2017optimal}
P.~Nespoli, D.~Papamartzivanos, F.~G. M{\'a}rmol, and G.~Kambourakis, ``Optimal
  countermeasures selection against cyber attacks: A comprehensive survey on
  reaction frameworks,'' \emph{IEEE Communications Surveys \& Tutorials},
  vol.~20, no.~2, pp. 1361--1396, 2017.

\bibitem{gordon2002economics}
L.~A. Gordon and M.~P. Loeb, ``The economics of information security
  investment,'' \emph{ACM Transactions on Information and System Security},
  vol.~5, no.~4, pp. 438--457, 2002.

\bibitem{chronopoulos2017options}
M.~Chronopoulos, E.~Panaousis, and J.~Grossklags, ``An options approach to
  cybersecurity investment,'' \emph{IEEE Access}, vol.~6, pp. 12\,175--12\,186,
  2017.

\bibitem{hasrouny2017vanet}
H.~Hasrouny, A.~E. Samhat, C.~Bassil, and A.~Laouiti, ``{VANET} security
  challenges and solutions: A survey,'' \emph{Vehicular Communications},
  vol.~7, pp. 7--20, 2017.

\bibitem{vermesan2011internet}
O.~Vermesan, P.~Friess, P.~Guillemin, S.~Gusmeroli, H.~Sundmaeker, A.~Bassi,
  I.~S. Jubert, M.~Mazura, M.~Harrison, M.~Eisenhauer \emph{et~al.}, ``Internet
  of things strategic research roadmap,'' \emph{Internet of things-global
  technological and societal trends}, vol.~1, no. 2011, pp. 9--52, 2011.

\bibitem{qu2015security}
F.~Qu, Z.~Wu, F.-Y. Wang, and W.~Cho, ``A security and privacy review of
  {VANETs},'' \emph{IEEE Transactions on Intelligent Transportation Systems},
  vol.~16, no.~6, pp. 2985--2996, 2015.

\bibitem{petit2014potential}
J.~Petit and S.~E. Shladover, ``Potential cyberattacks on automated vehicles,''
  \emph{IEEE Transactions on Intelligent transportation systems}, vol.~16,
  no.~2, pp. 546--556, 2014.

\bibitem{parkinson2017cyber}
S.~Parkinson, P.~Ward, K.~Wilson, and J.~Miller, ``Cyber threats facing
  autonomous and connected vehicles: Future challenges,'' \emph{IEEE
  transactions on intelligent transportation systems}, vol.~18, no.~11, pp.
  2898--2915, 2017.

\bibitem{zhang2014defending}
T.~Zhang, H.~Antunes, and S.~Aggarwal, ``Defending connected vehicles against
  malware: Challenges and a solution framework,'' \emph{IEEE Internet of Things
  journal}, vol.~1, no.~1, pp. 10--21, 2014.

\bibitem{miller2015remote}
C.~Miller and C.~Valasek, ``Remote exploitation of an unaltered passenger
  vehicle,'' \emph{Black Hat USA}, vol. 2015, p.~91, 2015.

\bibitem{li2021rted}
J.~Li, Z.~Xue, C.~Li, and M.~Liu, ``Rted-sd: A real-time edge detection scheme
  for sybil ddos in the internet of vehicles,'' \emph{IEEE Access}, vol.~9, pp.
  11\,296--11\,305, 2021.

\bibitem{singh2019machine}
P.~K. Singh, S.~Gupta, R.~Vashistha, S.~K. Nandi, and S.~Nandi, ``Machine
  learning based approach to detect position falsification attack in
  {VANETs},'' in \emph{International Conference on Security \& Privacy}.\hskip
  1em plus 0.5em minus 0.4em\relax Springer, 2019, pp. 166--178.

\bibitem{lo2007illusion}
N.-W. Lo and H.-C. Tsai, ``Illusion attack on {VANET} applications-a message
  plausibility problem,'' in \emph{2007 IEEE globecom workshops}.\hskip 1em
  plus 0.5em minus 0.4em\relax IEEE, 2007, pp. 1--8.

\bibitem{wang2020topology}
J.~Wang, Y.~Tan, J.~Liu, and Y.~Zhang, ``Topology poisoning attack in
  sdn-enabled vehicular edge network,'' \emph{IEEE Internet of Things Journal},
  vol.~7, no.~10, pp. 9563--9574, 2020.

\bibitem{carroll2011game}
T.~E. Carroll and D.~Grosu, ``A game theoretic investigation of deception in
  network security,'' \emph{Security and Communication Networks}, vol.~4,
  no.~10, pp. 1162--1172, 2011.

\bibitem{boumkheld2019honeypot}
N.~Boumkheld, S.~Panda, S.~Rass, and E.~Panaousis, ``Honeypot type selection
  games for smart grid networks,'' in \emph{International Conference on
  Decision and Game Theory for Security}.\hskip 1em plus 0.5em minus
  0.4em\relax Springer, 2019, pp. 85--96.

\bibitem{zhu2013game}
Q.~Zhu and T.~Ba{\c{s}}ar, ``Game-theoretic approach to feedback-driven
  multi-stage moving target defense,'' in \emph{International Conference on
  Decision and Game Theory for Security}.\hskip 1em plus 0.5em minus
  0.4em\relax Springer, 2013, pp. 246--263.

\bibitem{tian2021honeypot}
W.~Tian, M.~Du, X.~Ji, G.~Liu, Y.~Dai, and Z.~Han, ``Honeypot detection
  strategy against advanced persistent threats in industrial internet of
  things: A prospect theoretic game,'' \emph{IEEE Internet of Things Journal},
  2021.

\bibitem{you2020scalable}
J.~You, S.~Lv, L.~Zhao, M.~Niu, Z.~Shi, and L.~Sun, ``A scalable
  high-interaction physical honeypot framework for programmable logic
  controller,'' in \emph{2020 IEEE 92nd Vehicular Technology Conference
  (VTC2020-Fall)}.\hskip 1em plus 0.5em minus 0.4em\relax IEEE, 2020, pp. 1--5.

\bibitem{prathapani2009intelligent}
A.~Prathapani, L.~Santhanam, and D.~P. Agrawal, ``Intelligent honeypot agent
  for blackhole attack detection in wireless mesh networks,'' in \emph{2009
  IEEE 6th International Conference on Mobile Adhoc and Sensor Systems}.\hskip
  1em plus 0.5em minus 0.4em\relax IEEE, 2009, pp. 753--758.

\bibitem{babu2016novel}
M.~R. Babu and G.~Usha, ``A novel honeypot based detection and isolation
  approach (nhbadi) to detect and isolate black hole attacks in {MANET},''
  \emph{Wireless Personal Communications}, vol.~90, no.~2, pp. 831--845, 2016.

\bibitem{garg2007deception}
N.~Garg and D.~Grosu, ``Deception in honeynets: A game-theoretic analysis,'' in
  \emph{2007 IEEE SMC Information Assurance and Security Workshop}.\hskip 1em
  plus 0.5em minus 0.4em\relax IEEE, 2007, pp. 107--113.

\bibitem{cceker2016deception}
H.~{\c{C}}eker, J.~Zhuang, S.~Upadhyaya, Q.~D. La, and B.-H. Soong,
  ``Deception-based game theoretical approach to mitigate dos attacks,'' in
  \emph{International Conference on Decision and Game Theory for
  Security}.\hskip 1em plus 0.5em minus 0.4em\relax Springer, 2016, pp. 18--38.

\bibitem{kiekintveld2015game}
C.~Kiekintveld, V.~Lis{\`y}, and R.~P{\'\i}bil, ``Game-theoretic foundations
  for the strategic use of honeypots in network security,'' in \emph{Cyber
  warfare}.\hskip 1em plus 0.5em minus 0.4em\relax Springer, 2015, pp. 81--101.

\bibitem{zhuang2010modeling}
J.~Zhuang, V.~M. Bier, and O.~Alagoz, ``Modeling secrecy and deception in a
  multiple-period attacker--defender signaling game,'' \emph{European Journal
  of Operational Research}, vol. 203, no.~2, pp. 409--418, 2010.

\bibitem{durkota2015optimal}
K.~Durkota, V.~Lis{\`y}, B.~Bo{\v{s}}ansk{\`y}, and C.~Kiekintveld, ``Optimal
  network security hardening using attack graph games,'' in \emph{Twenty-Fourth
  International Joint Conference on Artificial Intelligence}, 2015.

\bibitem{nisioti2021data}
A.~Nisioti, G.~Loukas, A.~Laszka, and E.~Panaousis, ``Data-driven decision
  support for optimizing cyber forensic investigations,'' \emph{IEEE
  Transactions on Information Forensics and Security}, vol.~16, pp. 2397--2412,
  2021.

\bibitem{panda2020optimizing}
S.~Panda, E.~Panaousis, G.~Loukas, and C.~Laoudias, ``Optimizing investments in
  cyber hygiene for protecting healthcare users,'' \emph{From Lambda Calculus
  to Cybersecurity Through Program Analysis}, 2020.

\bibitem{fielder2018risk}
A.~Fielder, S.~K{\"o}nig, E.~Panaousis, S.~Schauer, and S.~Rass, ``Risk
  assessment uncertainties in cybersecurity investments,'' \emph{Games},
  vol.~9, no.~2, p.~34, 2018.

\bibitem{fan2017enabling}
W.~Fan, Z.~Du, D.~Fern{\'a}ndez, and V.~A. Villagra, ``Enabling an anatomic
  view to investigate honeypot systems: A survey,'' \emph{IEEE Systems
  Journal}, vol.~12, no.~4, pp. 3906--3919, 2017.

\bibitem{araujo2014patches}
F.~Araujo, K.~W. Hamlen, S.~Biedermann, and S.~Katzenbeisser, ``From patches to
  honey-patches: Lightweight attacker misdirection, deception, and
  disinformation,'' in \emph{Proceedings of the 2014 ACM SIGSAC conference on
  computer and communications security}, 2014, pp. 942--953.

\bibitem{avery2017ghost}
J.~Avery and E.~H. Spafford, ``Ghost patches: Fake patches for fake
  vulnerabilities,'' in \emph{IFIP International Conference on ICT Systems
  Security and Privacy Protection}.\hskip 1em plus 0.5em minus 0.4em\relax
  Springer, 2017, pp. 399--412.

\bibitem{fudenberg_game_1991}
D.~Fudenberg and J.~Tirole, \emph{Game {Theory}}.\hskip 1em plus 0.5em minus
  0.4em\relax London: MIT Press, 1991.

\bibitem{pitropakis2018enhanced}
N.~Pitropakis, E.~Panaousis, A.~Giannakoulias, G.~Kalpakis, R.~D. Rodriguez,
  and P.~Sarigiannidis, ``An enhanced cyber attack attribution framework,'' in
  \emph{International Conference on Trust and Privacy in Digital
  Business}.\hskip 1em plus 0.5em minus 0.4em\relax Springer, 2018, pp.
  213--228.

\bibitem{rass2017cost}
S.~Rass, S.~K{\"o}nig, and S.~Schauer, ``On the cost of game playing: How to
  control the expenses in mixed strategies,'' in \emph{International Conference
  on Decision and Game Theory for Security}.\hskip 1em plus 0.5em minus
  0.4em\relax Springer, 2017, pp. 494--505.

\bibitem{hausken2006returns}
K.~Hausken, ``Returns to information security investment: The effect of
  alternative information security breach functions on optimal investment and
  sensitivity to vulnerability,'' \emph{Information Systems Frontiers}, vol.~8,
  no.~5, pp. 338--349, 2006.

\bibitem{qassrawi2010deception}
M.~T. Qassrawi and Z.~Hongli, ``Deception methodology in virtual honeypots,''
  in \emph{2010 Second International Conference on Networks Security, Wireless
  Communications and Trusted Computing}, vol.~2.\hskip 1em plus 0.5em minus
  0.4em\relax IEEE, 2010, pp. 462--467.

\bibitem{ruiz2010optimization}
R.~Ruiz-Torrubiano, S.~Garc{\'\i}a-Moratilla, and A.~Su{\'a}rez, ``Optimization
  problems with cardinality constraints,'' in \emph{Computational Intelligence
  in Optimization}.\hskip 1em plus 0.5em minus 0.4em\relax Springer, 2010, pp.
  105--130.

\end{thebibliography}
\end{document}